\documentclass[12pt,preprint]{aastex}    
\usepackage{emulateapj5}             
\newcommand{\fuse}{{\it FUSE\/}}

\newcommand{\cii}{C\,{\sc ii}}       
\newcommand{\ciii}{C\,{\sc iii}}     
\newcommand{\civ}{C\,{\sc iv}}       
\newcommand{\siiv}{Si\,{\sc iv}}     
\newcommand{\siii}{Si\,{\sc ii}} 
\newcommand{\siv}{S\,{\sc iv}}       
\newcommand{\pv}{P\,{\sc v}}         
\newcommand{\nni}{N\,{\sc i}}        
\newcommand{\nii}{N\,{\sc ii}}       
\newcommand{\niv}{N\,{\sc iv}}       
\newcommand{\oi}{O\,{\sc i}}         
\newcommand{\ovi}{O\,{\sc vi}}       
\newcommand{\feii}{Fe\,{\sc ii}} 
\newcommand{\hei}{He\,{\sc i}}     
 
\received{2002 May 8}
\begin{document} 
 
\title{Synthetic High-Resolution Line Spectra of Star-Forming Galaxies 
Below 1200~\AA\altaffilmark{1}} 
 
\author{Carmelle Robert and Anne Pellerin} 
 
\affil{D\'epartement de physique, de g\'enie physique et d'optique, 
Universit\'e Laval and Observatoire du mont M\'egantic, 
Qu\'ebec, QC, G1K 7P4, Canada; carobert@phy.ulaval.ca, apelleri@phy.ulaval.ca} 
 
\author{Alessandra Aloisi} 
 
\affil{Department of Physics and Astronomy, Johns Hopkins University, 
3400 North Charles Street, Baltimore, MD 21218; aloisi@pha.jhu.edu} 
 
\author{Claus Leitherer} 
 
\affil{Space Telescope Science Institute\altaffilmark{2}, 
3700 San Martin Drive, Baltimore, MD 21218; leitherer@stsci.edu} 
 
\author{Charles Hoopes and Timothy M. Heckman} 
 
\affil{Department of Physics and Astronomy, Johns Hopkins University, 
3400 North Charles Street, Baltimore, MD 21218; 
choopes@pha.jhu.edu, heckman@pha.jhu.edu}

\altaffiltext{1}{Based on observations made with the NASA-CNES-CSA Far 
Ultraviolet Spectroscopic Explorer.  {\it FUSE} is operated for NASA by the 
Johns Hopkins University under NASA contract NAS5-32985.} 
 
\altaffiltext{2}{Operated by AURA, Inc., under NASA contract NAS5-26555.} 
 
\begin{abstract} 
We have generated a set of far-ultraviolet stellar libraries using
spectra of OB and Wolf-Rayet stars in the Galaxy and the Large and
Small Magellanic Cloud. The spectra were collected with the
{\em Far Ultraviolet Spectroscopic Explorer} and  cover
a wavelength range from 1003.1 to 1182.7~\AA\ at a resolution of 0.127~\AA.
The libraries extend from the earliest O- to late-O and early-B stars
for the Magellanic Cloud and Galactic libraries, respectively.
Attention is paid to the complex blending of stellar and interstellar
lines, which can be significant, especially in models using  Galactic
stars. The most severe contamination is due to molecular hydrogen.
Using a simple model for the H$_2$ line strength, we were able to remove
the molecular hydrogen lines in a subset of Magellanic Cloud stars.
Variations of the photospheric and wind features of \ciii\,$\lambda$1176,
\ovi\,$\lambda\lambda$1032,~1038, \pv\,$\lambda\lambda$1118,~1128,
and \siv\,$\lambda\lambda$1063,~1073,~1074 are discussed as a function of
temperature and luminosity class. The spectral libraries were implemented
into the LavalSB and Starburst99 packages and used to compute a standard
set of synthetic spectra of star-forming galaxies. Representative spectra
are presented for various initial mass functions and star formation histories.
The valid parameter space is confined to the youngest ages of less than
$\simeq$10~Myr for an instantaneous burst, prior to the age when 
incompleteness of spectral types in the libraries sets in.
For a continuous burst at solar metallicity, the parameter space is not
limited.  The suite of models is useful for interpreting the
restframe far-ultraviolet in local and high-redshift galaxies.

\end{abstract} 
 
 
\section{Introduction} 
 
Modeling of synthetic line profiles in stellar populations has traditionally
focused on the optical, near-infrared (IR), and satellite-ultraviolet (UV)
wavelength regions. Since most of the models are empirical, i.e., they utilize
observational stellar libraries, or they are theoretical with the need for
observational calibrations, their wavelength coverage is determined by the 
available body of observations.  Examples of this type of work
are in the compilation of Leitherer et al. (1996).
As a result of past and ongoing efforts, line-profile synthesis models
cover the wavelength range from 1200~\AA\ to 2.3~$\mu$m. The lower cut-off
is due to the transmission of the entrance windows and the optical
coatings of most UV spectrographs, such as {\it IUE} and those of {\it HST},
and the upper limit defines the regime where near-IR observations become
background-, rather than detector-limited.

Few strong features exist in the spectra of young stellar populations longward
of the K-band, and those that would in principle be observable are normally
totally diluted by galactic gas and dust emission longward of the L-band (e.g.,
Genzel \& Cesarsky 2000). Consequently there is little practical interest in
pushing the profile synthesis work to longer wavelengths. The situation is
different at short wavelengths. The spectral region of a young population
between 912 and 1200~\AA\ contains a rich absorption- and emission-line 
spectrum which provides important clues on the star-formation history 
(e.g., Gonz\'alez Delgado, Leitherer, \& Heckman 1997; Leitherer et al. 2002).
For instance, the strong resonance doublet of \ovi\,$\lambda$1032, 1038 is 
often stronger than the commonly used star-formation indicators 
\siiv\,$\lambda$1400 and \civ\,$\lambda$1550 (Robert, Leitherer, \& 
Heckman 1993).

Previously, the usefulness of the wavelength range 912~--~1200~\AA\ was 
limited due to the lack of observational data. Pioneering work by 
the Copernicus satellite (Rogerson et al. 1973) dramatically increased our 
understanding of stellar spectra but extragalactic sources were too faint 
for observation.  The Voyager~1 and 2 spacecraft (Broadfoot et al. 1977) 
had similar limitations.  More recently, the {\it ORFEUS} mission 
(Grewing et al. 1991) explored the wavelength region below 1200~\AA, but 
again, few faint objects could be observed due to brightness limitations. 
The Hopkins Ultraviolet Telescope ({\it HUT}; Davidsen 1993) was the first 
instrument sensitive enough to collect astrophysically useful spectra 
of faint galaxies in
the wavelength range below L$\alpha$. The short mission duration of two weeks
precluded the built-up of a significant database. Nevertheless, the spectra
of star-forming galaxies obtained with {\it HUT}
demonstrated the potential of this wavelength for our understanding of
star-forming galaxies (Leitherer et al. 2002).

Renewed interest in the far-UV region of star-forming galaxies is generated by
two major science drivers. First, the launch of the multi-year Far
Ultraviolet Spectroscopic Explorer (\fuse) mission (Moos et al. 1998) is
producing high-quality far-UV
spectra of numerous star-forming galaxies which await interpretation with
spectral synthesis modeling. Second, ground-based 10-m class telescopes can
observe the restframe far-UV spectra of star-forming galaxies at $z>3$ (Pettini
et al. 2001). These spectra reach down to the Lyman limit and achieve a
signal-to-noise rivaling that of UV spectra of their local counterparts.

As part of a \fuse\ Guaranteed Time Observer (GTO) program, spectra
of more than 200 hot, luminous stars in the Galaxy and in the Magellanic 
Clouds have been collected.  These data, while astrophysically useful in 
their own right, are well suited to extend currently existing spectral 
libraries longward of 1200~\AA\ down to the
Lyman limit. The individual spectra are discussed by Pellerin et al. (2002)
and Walborn et al. (2002a). In the present paper we describe the creation of
the far-UV spectral libraries, their implementation into existing evolutionary
synthesis codes, and a parameter study using a standard
grid of synthetic spectra produced with the new libraries.

\section{FUSE data} 
 
Our far-UV libraries contain spectra of OB and Wolf-Rayet (WR) stars collected
with \fuse\ mainly during its first cycle (1999 December to 2000 December).
The \fuse\ instrument provides redundant wavelength coverage by means of 8 
independent spectra, each covering a $\sim$100~\AA\ region between 905 and
1187~\AA\ at a resolving power of $\sim$20\,000. A detailed description of
\fuse\ can be found in Moos et al. (2000) and Sahnow et al. (2000).
Because of systematic differences in their spectral coverage and data
quality, a limited combination of these 8 spectra was chosen for the libraries.
The LiF2A segment (i.e. channel 2 and segment A of the LiF detector)
was selected to cover the long wavelength region from 1086 to 1183~\AA\ and
the SiC1A segment was adopted to cover the region between 1003 and 1092~\AA.
The SiC1A segment was chosen rather than the LiF1A segment (which
has greater sensitivity) because the LiF1A spectral coverage extends
only to about 1082~\AA\ (leaving a small gap in the spectral coverage).
Because of the increasing number of interstellar lines below 1000~\AA,
especially for the Galactic objects, which seriously  contaminate stellar
lines and the continuum, we simply decided not to extend
the libraries to these shorter wavelengths. The data used to
build the libraries were all collected with the LWRS aperture
($30^{\prime\prime} \times 30^{\prime\prime}$).
Typical exposure times are around 5\,000~s.

The spectra selected for the libraries were processed with the version 1.8.7
of the standard {\tt calfuse} calibration pipeline. This version did not flat
field the data or correct for astigmatism, but it provided flux and wavelength
calibrated spectra taking into account small thermal and electronic effects.
The flux calibration is accurate to better that 10\% and the wavelength scale
has a precision of about a resolution element ($\sim$15-20~km~s$^{-1}$)
because the LWRS aperture zero point is poorly known.
An independent reduction step was performed by us in order to correct for
to the heliocentric frame (this step was not properly done by the pipeline
version used). Furthermore, spectra of stars in the LMC and SMC were
deredshifted using a velocity of 278~km~s$^{-1}$ and 158~km~s$^{-1}$,
respectively (from the NASA/IPAC Extragalactic Database).

For the purpose of spectral synthesis, the spectrum of individual stars
included into the libraries has to be reddening free.
For many of the stars, this parameter is not known
with good accuracy. One way to avoid this problem is to normalize the
individual star spectra to a continuum equal to unity and to rescale,
within the synthesis codes (as described in \S4), the absolute flux
level using atmosphere models. The normalization was done by dividing each
individual segment spectrum by a function obtained from fitting a
low order (4 to 7) Legendre polynomial through continuum points that
avoid broad spectral features.  Next, the spectra from the two segments
were merged together using a simple average in the overlapping region.
Furthermore, in order to enhance the visibility of the stellar features,
the spectra were binned over 20 pixels.
The final spectra cover from 1003.1 to 1182.7~\AA\ with a resolution
of 0.127~\AA\, and have a typical S/N of 30 per resolution
element in the continuum in the LiF2A segment.

For about half the stars in the Magellanic Clouds we have been able to remove
the molecular hydrogen contribution associated to the Magellanic Cloud ISM
on the line of sight.  This operation was based on the work of Tumlinson
et al. (2002) who recently completed a survey of H$_2$ toward massive stars
in the LMC and SMC.
They measured the column densities of H$_2$ in the rotational levels
$J=1$ through $J=6$, as well as the Doppler broadening parameter $b$.
We created a simple model for each H$_2$ absorption line using the
column density and $b$-value to calculate the optical depth at the center
of the line, and then approximating the optical depth per unit velocity as
a Gaussian function with $\sigma=b/[2^{(1/2)}]$. The wavelengths and 
$f$-values for the H$_2$ lines were taken from Abgrall \& Roueff (1989). 
The resulting spectrum ($I_{\rm{v}}=e^{-\tau_{\rm{v}}}$) was then 
convolved with an approximate \fuse\ line
spread function consisting of two Gaussian components (for the SiC1A
channel we used a narrow component with FWHM=10 pixels, and a broad
but weaker component with FWHM=20 pixels, and 8 and 19 pixels for
LiF2A). The stellar spectrum was then divided by the convolved model spectrum.
The use of the Gaussian function to approximate the optical
depth profile results in an underprediction of the strengths of
saturated lines (in which case a Voigt function would work better),
but this works to our advantage by avoiding the complication of
correcting the saturated lines. These lines have intensity very close
to zero, so dividing by a model that approaches zero would produce
huge noise spikes in the line, instead of removing it. Our model
predicts much weaker lines, essentially leaving the saturated lines
uncorrected, while removing the unsaturated lines.  There is
no survey so far which address the H$_2$ in the Milky Way along
any star sight lines, so the Galactic H$_2$ absorption remains in all
stellar spectra in our libraries.  Figure~\ref{fig0} shows the H$_2$ line 
identification for a typical Galactic, LMC, and SMC star. 
For the Magellanic Cloud objects, both spectrum, before and after the 
removal of the Cloud H$_2$ line contribution, is plotted.
The study of the far-UV spectral features (stellar and/or
interstellar) of a low metallicity galaxy will be
simplified when comparing its far-UV spectrum with a synthetic model
based on a library which is as free as possible of intrinsic
H$_2$ contamination.

For the Magellanic Cloud spectra only we have removed the
strongest scattered \hei\ solar emission, at 1168.7~\AA. This feature is
due to the second order of the gratings and its strength depends
on the spacecraft orientation and on the exposure time.

The library stars have been selected from the large pool of \fuse\
hot star data using two criteria: the quality of the observation
(i.e. high S/N) and the quality of the spectral classification. Objects
which show peculiar spectral features compared with other stars within
the same spectral type group were rejected. Often the classification
of these objects (essentially only Galactic stars) is based on old 
photographic plates, and therefore merits a revision based on today's 
conventions.  The \fuse\ atlas of hot stars in the Galaxy and Magellanic 
Clouds by Pellerin et al. (2002) and Walborn et al. (2002a), respectively, 
has been used as a reference for the stellar line characteristics as a 
function of the spectral type and luminosity class.

With these criteria, 155 stars in the Galaxy covering the spectral
types from O3 to B3 have been selected. A total of 41 stars in the LMC
and 32 stars in the SMC cover the spectral types from O3 to B0 with
some gaps for the hotter types.  The limit for the late type stars
was imposed by the fewer stars available in the present \fuse\ archive.
All five luminosity classes (from dwarfs to supergiants) are represented,
with more gaps in the coverage for the subgiants and bright giants.
WR stars of subtype WN and WC are also included: 13 for the Galaxy,
and 9 and 2 for the Large and Small Magellanic Cloud, respectively.
Tables~1 and 2 present the fundamentals properties of the selected objects
for the far-UV solar-metallicity and the Magellanic Cloud libraries,
respectively.  In these tables, successive columns contain the star name,
J2000 coordinates, spectral classification, source of the classification,
apparent visual magnitude, color excess, and \fuse\ identification
for the data set as specified in the Multi-Mission Archive at the Space
Telescope Science Institute.

\section{Generation of the libraries} 
 
In order to reflect the differences in metallicity of the stellar lines
and the importance of the H$_2$ contribution, four libraries have been
created: a solar-metallicity library, an intermediate metallicity library 
using LMC stars, an SMC library at lower metallicity,
and a LMC+SMC library at an average metallicity between the two Clouds and
combining a relatively well balanced number of stars from both Clouds.
The individual LMC and SMC libraries are useful to study
young star forming regions using the LavalSB code (Dionne \& Robert 2002)
where stellar evolutionary tracks at the metallicity of the LMC and SMC 
have been interpolated.  In order to keep a maximum number of stars in 
these libraries the removal of the H$_2$ contribution from the 
Magellanic Cloud on the line of sight was not performed (i.e. about 
only half of these stars have known H$_2$ column densities so far). 
The LMC+SMC library was created for the Starburst99 code 
(Leitherer et al. 1999) where evolutionary tracks at the intermediate 
metallicity of the Clouds are available. Because it includes more stars,
it was possible to restrict this library to objects for which the H$_2$
contribution on the line of sight from the corresponding Magellanic Cloud
is known. The LMC+SMC library, therefore, has only spectra of stars where the
H$_2$ contribution from the Magellanic Clouds was removed
following the technique described in \S2.
Ultraviolet libraries (LMC, SMC and LMC+SMC) in the 1200~--~1600~\AA\ range
are already included in the synthesis codes, which contain about the same
number of stars than the far-UV libraries and where more that half of 
these stars are found in both wavelength ranges (e.g. Leitherer et al. 2001).

While creating the libraries, we averaged the spectra of stars with
the same spectral type in order to increase the S/N of the spectra
and to minimize the impact of spectral misclassification.
Because of the fewer stars available in some spectral type groups and
their lower S/N, we occasionally combined two successive groups.
The format of the libraries is a two
dimensional grid. We use 21 temperature classes, from O3, O3.5, O4,
O4.5, ..., to B3, for the Galactic stars or 15 classes, between O3 and B0,
for the Magellanic Cloud stars. We also use five luminosity classes:
V, IV, III, II, and I. Four groups represent the WR stars: WNE (which
includes WN3 to WN6 stars), WNL (WN7 to WN11), WCE (WC4 to WC5), and
WCL (WC6 to WC9). This grid is dictated by the interpolation
scheme in the Starburst99 and LavalSB codes.

Tables~3 and 4 give a list of the stars included in each spectral type
group and summarize the rules adopted for the extrapolation (in temperature,
luminosity, or metallicity) of important missing groups.
Groups not listed in these tables have been created from an
interpolation in temperature. Specification of an LiF or SiC segment
for an individual star indicates that the counterpart segment was rejected
because of a low signal (possibly due to misalignment problems).

For the solar-metallicity library (Table~3), most of the O subgiants
and bright giants (from O3 to O8) and the early O3 to O4.5 giants have been
interpolated between luminosity classes. It seems reasonable to adopt this
interpolation based on the smooth variation of the {\it HST}/UV stellar lines
between 1120 and 1600~\AA\ with luminosity classes
(e.g., see Walborn, Nichols-Bohlin, \& Panek 1985).
The missing B3I group is simply a copy of the B2I star, where at this
surface temperature few changes in the spectral lines for nearby types
are expected. As previously mentioned, the omitted groups in the
solar-metallicity library table have been created by means of an 
interpolation in temperature (e.g. the spectrum for the O3.5V group 
is an average between O3V and O4V, etc).

The individual libraries for the LMC and SMC contain fewer
stars compared to the solar-metallicity library and therefore required a more
drastic extrapolation for the missing groups. As indicated in Table~4,
in the case of the LMC library, an extrapolation in metallicity between
the solar-metallicity library and the SMC library was done for the O3V and
O9 to B0 dwarfs and giants groups.
A comparison with the spectrum of nearby groups
(i.e. the O3V versus O4V to O6V, and B0V, B0III versus B0I) shows a smooth
transition. In the case of the SMC library, the O3I to O5.5I are a
copy of the O3III group for the SMC and the O4I to O5.5I groups in the LMC
library. This extrapolation gives a smooth transition of the line properties,
with a large uncertainty for the absolute strength of the stellar lines
in the SMC hot supergiants as described in more detail in the following
paragraphs.  Fewer WR stars in the SMC have been observed with \fuse\ 
during the first cycle. The missing WR groups are therefore a copy of 
the LMC and solar-metallicity library groups.  In a standard stellar 
population, the WR contribution to the global stellar line
profiles occurs during a brief evolutionary phase (a few 10$^6$~yr),
and is not very strong because of the relatively small number of WR stars.
This minimizes the effect of the metallicity extrapolation among the 
WR stars for the synthesis.

The LMC+SMC library offers (albeit with a restricted number of stars) a good
coverage of all the spectral type groups with a reasonably well-balanced
distribution of the LMC and SMC spectra among the groups.
As listed in Table~4, there are 27 stars from the LMC and 19 from the
SMC which are fairly well distributed between the OB spectral type groups.
For this library every spectrum of each individual star has
had the H$_2$ contribution from the Magellanic Cloud removed (see \S2).
Since not all Magellanic Cloud stars have known H$_2$ column densities,
there are stars in the
SMC or LMC libraries that are not included in the LMC+SMC library.

In the following paragraphs we summarize the variation with spectral
type, luminosity class, and metallicity of the most interesting stellar
indicators in the far-UV spectral range based on our libraries. More details
about the stellar lines (and identification of interstellar lines)
may be found in the atlas of Pellerin et al. (2002) for the Galactic stars
and Walborn et al. (2002a) for the Magellanic Cloud stars.
Figures~\ref{fig1} to \ref{fig3} show these stellar indicators
grouped by luminosity class. These figures, used here to describe the stellar
lines, are also very interesting as they provide a quick feeling
about the composite spectrum of a young stellar population
dominated by main sequence hot stars (before $\sim$3~Myr) or more
evolved hot stars.

\ \ 
 
\centerline{\it \ciii\,$\lambda$1176} 
 
In main sequence OB stars, the \ciii\ multiplet centered at 1175.6 \AA\
displays a strong photospheric absorption, independent of the metallicity
(see Fig.~\ref{fig1}). The absorption slowly increases in strength with
decreasing temperature down to about B1-B3 dwarfs (and then becomes weaker
in cooler B stars).

For the giant stars, the \ciii\ multiplet is dependent on
the metallicity with a stronger wind P~Cygni profile for Galactic stars
(compare Fig.~\ref{fig1} to \ref{fig3}).
More specifically, Galactic O5 to O9.5 giants show a broad
absorption due to a stellar wind, superimposed on a distinct photospheric
absorption line. For the O3 to O4 giants only the photospheric contribution
is seen (as in the case of dwarfs of the same temperature class).
In giant stars cooler than B0-B1, the wind signature is absent and strong
photospheric absorption is seen with a similar strength as in the
corresponding dwarfs.
In lower metallicity objects, the wind profile seen for the mid and late
O giants in the Galaxy is barely detectable:
a weak indication is found only for the LMC O8-O9 giants, and no
blue shifted absorption or emission related to a stellar wind is seen
for the SMC giants. In these low metallicity environments,
the \ciii\ multiplet is dominated by the photospheric contribution.

In Galactic supergiants, a \ciii\ wind profile starts to appear at O4
and becomes the primary contribution to the line at
O6 to O9 (the photospheric absorption is completely lost in the broad
absorption component of the P~Cygni wind profile; see Fig.~\ref{fig3}).
A weak photospheric \ciii\ multiplet
is seen in O3I, and supergiants of subtype B0 and cooler show strong
photospheric absorptions. LMC supergiants display similar \ciii\ profiles
to the Galactic stars, except that the wind signature is relatively
weaker with less emission and narrower absorption. SMC supergiants
also show a wind signature in their \ciii\ multiplet between spectral
types O5 to O9 which is weaker than for LMC stars of the same temperature
class. The LMC+SMC library
represents very well the average \ciii\ profile behavior with metallicity
of the individual LMC and SMC libraries.

\ \ 
 
\centerline {\it \ovi\,$\lambda\lambda$1032, 1038} 
 
The \ovi\ doublet at 1031.9 and 1037.6\AA\ correlates with
temperature class and metallicity but not so much for the luminosity class
(refer to Fig.~\ref{fig1} to \ref{fig3}).
In the Milky Way, the \ovi\ doublet of early O stars shows a single P~Cygni
wind profile with a broad absorption trough blended with the saturated
interstellar L$\beta$ line, and a redshifted emission peak near 1040 \AA.
In late O type stars, two distinct blueshifted absorption components are
often visible. However, the emission component (if present) is removed by
the absorption due to both the \ovi\ member at 1038~\AA\ and the
interstellar \cii\,$\lambda$1036.3, 1037.0 lines.
In about B1.5 type stars a broad photospheric L$\beta$ line is starting
to develop, which gets stronger in cooler B stars.
There is no significant distinction between the \ovi\ profiles for dwarfs,
giants, and supergiants of the same temperature class (an exception being
the width of the absorption component, which seems narrower in a few
hotter supergiants).

A clear variation of the \ovi\ doublet is observed with metallicity.
At lower metallicity, the \ovi\ feature changes with spectral type
in a similar fashion as in the case of the Galactic stars.
However, the strength of the P~Cygni feature
(mainly the width of the absorption component) decreases with
decreasing metallicity.
In the SMC stars, the \ovi\ wind signatures are very weak. The presence
of the wind is still suspected in these stars mainly because the red wing of
the L$\beta$ line is asymmetrical.

\ \ 
 
\centerline {\it \pv\,$\lambda\lambda$1118, 1128} 
 
All dwarf OB stars from the SMC, LMC and solar-metallicity library show 
a \pv\ doublet at 1118.0 and 1128.0~\AA\ which is a photospheric 
absorption feature (Fig.~\ref{fig1}). The \pv\ component at 1118~\AA\
becomes slowly weaker with later spectral type and shows a significant 
drop in strength for B1-B3 stars. The \pv\ component at 1128\AA\ is 
blended with the longer-wavelength member of the 
\siiv\,$\lambda\lambda$1122.5, 1128.3 doublet. The \siiv\ and \pv\ 
doublets have opposite behavior when going to later spectral type. 
Therefore, the 1128~\AA\ blend remains relatively independent of spectral 
type.

For some Galactic giants and supergiants, wind profiles of \pv\ are seen,
with the supergiants showing the strongest profiles. In the O3 groups,
a weak and global P~Cygni line is found superimposed on the photospheric
component. The absorption is broad, reaching the wavelength
$\sim$1105-1110~\AA\ where strong H$_2$ and \feii\ features are observed
(Fig.~\ref{fig2} and \ref{fig3}).
Only one very weak emission peak is seen around 1130~\AA. Between the
temperature groups O3.5 and O8, two distinct \pv\ wind profiles are found
(with a maximum strength in the O6 groups). With later spectral type,
the wind absorption components get narrower. In the case of late O and
early B giants and supergiants, the \pv\ doublet is photospheric with 
a similar behavior as in dwarfs.

At lower metallicity, the contribution from the stellar wind to the
\pv\ doublet is considerably reduced. Among the Magellanic giant stars,
only the O4 and O5 groups from the LMC library show signs of a stellar
wind. In these spectra, weak emission is barely noticeable while the
absorption component is difficult to identify because of the
relatively strong interstellar lines.
A \pv\ wind signature is barely found in the giant O stars from the
combined LMC+SMC library (see Fig.~\ref{fig2}).
In the case of Magellanic supergiant stars, wind profiles of
\pv\ are clearly seen. As for the Galactic stars, a weak P~Cygni profile
is seen in O3I from the LMC. The O3.5 to O8 supergiants from the LMC display
two emission peaks. Nevertheless, the width of the wind absorption component
is smaller in the LMC supergiants compared to the Galactic stars.
The SMC early O supergiant groups display a smooth transition in line
properties from O3 to O7, with a consistent behavior in metallicity
(the hotter SMC supergiant groups being the result of an extrapolation
in metallicity and luminosity class).

\ \ 
 
\centerline {\it \siv\,$\lambda\lambda$1063, 1073, 1074} 
 
The \siv\ multiplet at 1062.7, 1073.0, and 1073.5~\AA\ displays similar
variations with spectral type, luminosity class, and metallicity as the
\ciii\,$\lambda$1176 feature. The component of the \siv\ multiplet at
1063~\AA\ is difficult to identify because it is blended with strong
features from H$_2$ and other interstellar elements (see Fig.~\ref{fig1}
to \ref{fig3}). For dwarf stars at
all metallicities, the components at 1073 -- 1074~\AA\ are a weak
photospheric feature in the early O groups. It increases in strength in
late O- early B dwarfs.

For giants stars, a weak wind signature is found only for the Galactic
O5 to O7 groups. It shows two broad blueshifted absorption features and
a weak emission peak for the red component.

Clear wind profiles are seen for the supergiants at all metallicities.
As for the \ciii\,$\lambda$1176 multiplet,
blueshifted absorption starts to appear in O4I. The emission component
of the wind profiles develops a maximum strength around type O6I,
and disappears quite suddenly in B0I.
The wind profiles are weaker for the lower metallicity supergiants,
but at lower metallicity, two emission peaks are seen
due to the 1063 and 1073 -- 1074~\AA\ lines. For the corresponding
Galactic supergiants the wind profiles are broader and the emission
portion of the 1063~\AA\ line is superimposed on the absorption part
of the 1073 -- 1074~\AA\ components.

\ \ 
 
\centerline {\it Other stellar lines} 
 
There are other lines that, while weaker than the lines discussed above
are also fairly good indicators of the stellar content, mainly because
they show some dependence on temperature. Among them, the absorption
line at 1169~\AA, often superimposed on the wind profiles
of \ciii\,$\lambda$1176 in the case of giants and supergiants, is possibly
a blend of \civ\,$\lambda$1168.9, 1169.0 with \niv\,$\lambda\lambda$1168.6,
1169.1, 1169.5. In contrast to \ciii\,$\lambda$1176, it reaches a maximum
strength in early O dwarfs and disappears in B dwarfs. The photospheric
multiplet \siv\,$\lambda$1099 increases in strength with increasing
temperature in a similar way to \siv\,$\lambda\lambda$1063,~1073,~1074
On the other hand, it does not show a wind signature for the supergiants.
 
\section{Implementation of the libraries in the population synthesis codes} 
 
We implemented the far-UV libraries in the population synthesis codes
Starburst99 and LavalSB. Starburst99 is a publicly available multi-use
application developed
at Space Telescope Science Institute to generate observables of young stellar
populations (Leitherer et al. 1999). LavalSB was derived from Starburst99 and
implemented at the Universit\'e Laval by one of us (Dionne \& Robert 2002).
The subroutine {\tt fusesyn} created for the synthesis of the far-UV spectrum
is described below. The two codes are virtually identical with respect to the
computation of UV line spectra. The main difference between the codes is the
metallicity of the stellar evolutionary tracks available.

The new subroutine {\tt fusesyn} was developed from an existing subroutine
called {\tt ovi} in Starburst99 to calculate the line spectrum around \ovi.
The old subroutine made use of a library of Copernicus spectra of OB stars to
calculate the line spectrum between 1016 and 1061~\AA\ (Gonz\'alez Delgado
et al. 1997). {\tt fusesyn} is a generalization of {\tt ovi} to different
metallicities and a broader wavelength range.

The computational technique is the same as in our previous work (Robert,
Leitherer, \& Heckman 1993; Leitherer, Robert, \& Heckman 1995; de Mello,
Leitherer, \& Heckman 2000; Leitherer et al. 2001).
Stars form according to a specified star-formation history and initial mass
function (IMF) and follow predefined tracks in the Hertzsprung-Russell
diagram (HRD) according to the evolution models of the Geneva group (Schaller
et al. 1992; Schaerer et al. 1993a,b; Charbonnel et al. 1993; Meynet et al.
1994).

The assignment of the individual library spectra for each star in the HRD,
following $T_{\rm{eff}}$ and $\log g$ of the evolutionary models,
is done using the empirical calibration of Schmidt-Kaler's (1982).
While more modern calibrations for O stars are available (e.g., Vacca, 
Garmany, \& Shull 1996), they are all within the random and systematic 
errors of
the earlier Schmidt-Kaler calibration. A discussion of this point was given
by Leitherer et al. (2001).  Martins, Schaerer, \& Hillier (2002) revisited
the effective temperature scale of O stars and favor the Schmidt-Kaler over
the Vacca et al. scale.

The library stars have a normalized continuum (as explained in \S2, because
of the uncertainty of the reddening for many stars), therefore spectral
types not included
in the libraries (down to late-type stars) have fluxes of unity at all
wavelengths. We will consider the consequences of this in \S5. The absolute
fluxes for each of these stars are obtained from model atmospheres.
For stars in the pre-WR phase, we scale the spectra using a stellar atmosphere
model of given $T_{\rm{eff}}$ and $\log g$ from the compilation of
Lejeune, Cuisinier \& Buser (1997) in the case of Starburst99 and
Kurucz (1992) in the case of LavalSB.  WR stars are modeled with the
spherically extended models of Schmutz, Leitherer, \& Gruenwald (1992) in
both codes. In practice, WR stars do not contribute significantly to the
wavelength range 912~--~1200~\AA\ for a standard population, and the details
of the computation do not matter. These atmosphere models are line-free, i.e.
the spectral lines were removed by fitting a multi-order spline to 
line-free regions.

At each time step, the spectra for each individual star present in the
specified stellar population are superposed by computing a stellar-number-
and flux-weighted average.  The line-free atmosphere set is then multiplied 
by the normalized fluxes of the library spectra, thereby ``adding'' spectra 
at \fuse\ resolution to the luminosity calibrated atmospheres.  The output 
product of the {\tt fusesyn} subroutine are synthetic far-UV spectra
between 1003.1 and 1182.7~\AA\ at 0.127~\AA\ resolution for O-star dominated
populations of arbitrary age, star-formation histories and IMF.
The stellar population spectra are computed both in normalized and in
absolute luminosity units.  In Starburst99, the metallicities are either 
solar or the average of the LMC+SMC. In LavalSB, they are either solar, 
LMC or SMC.  In both codes, the stellar evolution models and the library 
are consistent in their chemical composition.

\section{Synthetic galaxy spectra} 
 
We have computed a series of synthetic spectra for star-forming regions with
different IMFs and star formation histories. We will mainly concentrate here
on the discussion of models calculated with solar and average subsolar
(i.e., $\sim$~1/4 solar) metallicities using the evolutionary tracks at the
appropriate metallicity and the corresponding spectral library (i.e.,
solar-metallicity and LMC+SMC libraries).
We have considered two types of star-formation rate: an instantaneous burst,
where the star formation occurs all at the same time (i.e., at age zero);
and a continuous burst, where new stars are formed continuously, i.e., where
new stars appear, following the IMF prescription, at each time step (e.g.,
1~M$_\odot$~yr$^{-1}$).  We have varied the IMF in order
to modify the relative contribution of very massive stars with respect to
lower mass stars and to study the effects of this variation on the far-UV
spectra. A standard model is defined with a Salpeter IMF slope 
$\alpha$~=~2.35 and a lower and upper cut-off mass of 1 and 100~M$_{\odot}$, 
respectively.  We have also experimented with a steeper Miller-Scalo IMF 
($\alpha$~=~3.3) and a much flatter slope of $\alpha$~=~1.5, between 
the same mass range.  Finally truncated Salpeter IMFs with the upper 
cut-off mass of 30 and 50~M$_{\odot}$ have been considered.

The stellar library at solar metallicity extends until the B3 spectral 
subtype, while only spectral groups up to B0 are available for the subsolar
metallicities (LMC/SMC). This can affect the computed synthetic spectra of
an evolving stellar
population, depending on the IMF and/or age. It is important to stress here
that the stellar wind lines are from the most massive ($> 40$~M$_{\odot}$)
O stars and early B supergiants while all OB stars contribute significantly
to the far-UV continuum. If there were a non-degenerate mass-age relation,
as it applies to main-sequence stars, then a validity range for the age of a
stellar population could be easily assessed from the following reasoning.
The lifetime of the least massive O stars is around 10~Myr at both solar
and LMC/SMC metallicities (Schaller et al. 1992).
In the case of a standard {\it instantaneous} stellar population at subsolar
metallicity, with a spectral library containing up to B0 stars,
the synthetic spectra should be accurately reproduced until $\sim$~10~Myr. 
This validity age would be stretched up to $\sim$~15~Myr in the 
solar-metallicity
models, taking into account that later spectral groups until B3 are present.
In the evolution of real stellar populations, however, such mass-age relation
becomes degenerate due to the appearance of supergiants: e.g., B supergiants
evolving from the hottest O stars appear while there is still late 
O main-sequence stars (see Fig.~\ref{nbstar}).
Since supergiants of spectral types later than B3/B0 are not included
in our solar/subsolar-metallicity libraries, the age of incompleteness sets
in a few Myr earlier than expected. For ages greater than this completeness 
limit, only the theoretical atmosphere models will be available to produce 
at least a correct absolute value of the far-UV flux. Spectral features will
not be available anymore to constrain the age. In summary, before and around 
the completeness limit for the synthetic integrated spectrum of a stellar 
population, stellar lines from the library spectra will be diluted by the 
featureless continuum from cooler stars approximated by theoretical atmosphere
models. The lack of photospheric lines from late B stars in our models will 
make the wind emission peak stronger than observed, and the red side of 
the wide blue-shifted absorption
component weaker. Nevertheless, this leaves the bluest part of the wind
lines still very useful to constrain the age of a young stellar population.
After the age corresponding to the completeness limit, the stellar features,
mostly photospheric then, will be weaker than observed.

If the star-formation regime is {\it continuous}, the far-UV spectrum 
reaches a near-equilibrium after about 10~Myr. For a continuous burst, 
synthetic spectra
are always a mix of observed spectra for massive stars and theoretical
featureless atmosphere models for less massive stars. This implies that the
massive star line profiles predicted by a synthetic model would be correct
for all the ages considered, if B supergiants (the descendant of massive
main-sequence stars) could be neglected. Therefore, subsolar-metallicity
models for a
standard burst with a continuous star-forming rate are again limited to ages
below $\sim$~10~Myr (i.e. the subsolar-metallicity libraries stop at B0).
In the case of models at solar metallicity, the inclusion of B3 stars in the
library allow us to trust the continuous models to very old ages 
($>$~100~Myr).  Again, using stellar featureless atmosphere models for 
later type stars implies a good determination of the far-UV continuum 
level in the synthetic spectra for continuous star-forming regions. 
However, the absence of the photospheric absorption
component will affect the synthetic stellar lines, leaving only the bluest
part of the wind lines as useful age indicator.

\subsection{Instantaneous bursts} 
 
A time series for an instantaneous starburst between 0 and 10~Myr with
a standard Salpeter IMF and solar metallicity is considered in
Figure~\ref{inst.sol.Salp.100}. The corresponding model at subsolar 
metallicity is presented in Figure~\ref{inst.sub.Salp.100}.
These figures present model spectra which have been normalized to
a continuum equal to unit. This allow a better comparison of the
spectral features (but see \S5.6 for a discussion of the far-UV absolute
luminosity and continuum slope). The standard model, independent
of the metallicity used, clearly shows the general behavior of the wind lines
of \ovi\,$\lambda\lambda$1032,~1038, \siv\,$\lambda\lambda$1063,~1073,~1074,
\pv\,$\lambda\lambda$1118,~1128, and \ciii\,$\lambda$1176. The wind profiles
gradually strengthen from a main-sequence dominated stellar population of O
stars at 0-1~Myr to a population of luminous O supergiants at 3-4~Myr. After
this age, the emission component of the P~Cygni profile decreases until the
line spectrum is due to a population of early B stars at 10~Myr. 
The absorption component of stellar wind lines is not so strongly 
related to the presence of
massive OB stars due to the contribution of the interstellar medium
(mainly H$_2$) to their strength. The only exception is \ciii\,$\lambda$1176
at longer wavelength where there is no interstellar contribution.
At $\sim$~7~Myr the spectrum at subsolar metallicity already starts 
to show the dilution effects we were mentioning at the beginning of \S5.

The most important absorption features seen in Figures~\ref{inst.sol.Salp.100}
and \ref{inst.sub.Salp.100} can originate in stellar photospheres, in winds, or
in the ISM. In general, high excitation lines have a photospheric origin; only
in very rare cases they can originate in very dense stellar winds. One of the
strongest lines in the synthetic far-UV spectra is \ciii\,$\lambda$1176: it is
mostly photospheric with some wind contribution. It is one of the best
diagnostics for the age of a young burst. The \civ/\niv\ photospheric line
at 1169~\AA\ is also a fairly good diagnostic of the age, as it is present
and relatively strong in hot O stars only.
The high excitation line of \pv\
at 1118~\AA, with a mix of wind and photospheric contributions, can also be
very useful to distinguish between a very young and a $\sim$~8~Myr stellar
population (the second component of the doublet at 1128~\AA~ is blended
with \siiv\,$\lambda\lambda$1122,~1128). If detected, \pv\ shows a stronger
photospheric component in a younger burst and a weaker feature in a B star
dominated population. The standard model shows also that the high excitation
lines of \siiv\,$\lambda\lambda$1122,~1128 are photospheric with no strong
dependence on the temperature in O and B stars. These absorption lines
are nevertheless the only reasonably strong excited lines, after
\ciii\,$\lambda$1176, that we can see in the far-UV spectra of galaxies and
that, if detected, directly
assess the presence of massive OB stars in their stellar population. All other
spectral lines are resonance transitions. If the far-UV flux is dominated by O
stars, highly ionized species like \ovi\,$\lambda\lambda$1032,~1038 come from
stellar winds with some interstellar contribution.
The doublet of \ovi\ is blended with L$\beta$~$\lambda$1026 (which has a
mixed contribution from both the ISM and the photosphere in later B stars),
interstellar \cii\,$\lambda\lambda$1036.3, 1037.0, and H$_2$ lines from our
own Galaxy. Nonetheless, its nearly universal strength in all O stars and its
remarkable P~Cygni profile makes it a good diagnostic of young ($<~5$~Myr)
stellar populations. On the other hand, low-ionization species are from the
ISM: some examples of strong purely interstellar
lines are \siii\,$\lambda$1020.7, \cii\,$\lambda\lambda$1036.3, 1037.0,
\oi\,$\lambda$1039.2,
\nii\,$\lambda$1084.0 (plus its excited levels), \feii\ at 1063.2, 1096.9, and
1144.9~\AA, and \nni\,$\lambda\lambda$1134.2, 1134.4, 1135.0 (see Table 3 in
Pellerin et al. 2002 for a more complete list of interstellar lines).

All the solar-metallicity models show strong absorption lines of molecular
hydrogen in the wavelength range $<$~1110~\AA~ due to the strong contribution
from the Milky Way to the spectra in the solar-metallicity library. Absorption
lines from the Lyman and Werner rotational-vibrational bands of H$_2$,
arising from rotational levels J~=~0~--~7 in the ground vibrational state
are clearly present in the solar-metallicity models.  Stellar features at
wavelengths shorter than 1110~\AA~ in these models must be used with caution
when constraining the age of the stellar population. The Pellerin et al. (2002)
atlas of Galactic OB stars observed with \fuse\ presents the contribution
of interstellar H$_2$ and H~I in the far-UV band for different column
densities: their Figures~19 and 20 can be easily used to provide approximate
line identifications and estimates of the contamination effects by H$_2$ in
our synthetic spectra. On the other hand, subsolar-metallicity
models created with the LMC+SMC library do not
show such strong H$_2$ features: the spectra of the Magellanic stars in the
LMC+SMC library have been corrected for the Magellanic H$_2$ component
(see \S2), and the contribution from the Milky Way is
negligible due to the relatively high galactic latitude of these two galaxies.
The only residual structure related to molecular hydrogen in this library is
the small depression in the modeled spectra around 1050~\AA, a region that
is anyway free of stellar features.

\subsection{Metallicity effects} 

The two instantaneous models presented in Figures~\ref{inst.sol.Salp.100}
and \ref{inst.sub.Salp.100} differ from each other in their metallicity.
This implies that at the initial time (i.e., 0~Myr), for a zero-age
main-sequence stellar population, most of the photospheric and wind lines
are weaker at lower metallicity. At ages up to 4-5~Myr, when the winds
are important, the emission in the P~Cygni profile of \ciii\,$\lambda$1176
is weaker due to the lower
mass-loss rates at lower metallicity, analogously to \civ\,$\lambda$1550 at
longer wavelengths. The absorption component of \ciii\,$\lambda$1176
is narrower and weaker as well at lower metal content, like almost all
photospheric lines from O stars that scale with metal content.

The second-strongest stellar lines in strength which can be directly compared
in the two models at different metallicities are those in the
\pv\,$\lambda\lambda$1118, 1128 doublet. In this case the trend seems reversed:
the emission is stronger at lower metal content, particularly for the bluer
component. The absorption components of the doublet instead scale regularly
with metallicity, being narrower and weaker at subsolar metallicity.
A priori one would expect the emission
component in the P~Cygni profiles to be weaker at lower metallicity, but some
physical explanations can be found. A first effect is that the emission of
the \pv\ line at 1118~\AA~ appears stronger at lower metallicity because at
solar metallicity it has been weakened. Indeed at higher metallicity, the
absorption component of a wind P~Cygni profile is usually wider and more
extended towards shorter wavelengths. Therefore the absorption from \pv\
at 1128~\AA\ in the solar-metallicity models may extend
to the blue component of the doublet, and eats part of its emission. Stellar
winds with a velocity of the order of $\sim$ 2700~km~s$^{-1}$ (a quite normal
observed value) would be required by the wavelength separation of the doublet
components for this effect. The subsolar-metallicity
models have lower wind velocities,
and the blueshifted absorption due to the \pv\ line at 1128~\AA\ does not
reach the emission component of the \pv\ line at 1118~\AA. A second effect is
related to the competing ionization/recombination processes in stellar winds
at various metallicities. There are some empirical and theoretical
examples of stellar wind lines at longer UV wavelengths that show a higher
emission component at lower metallicity. Although the wind emission part of the
\pv\ line at 1128\AA\ does not appear stronger at lower metallicity, we are
still considering this effect because of the presence of interstellar
lines in the emission
region around 1130\AA\, the uncertainty of the normalization of the continuum
and the blend of \pv\ at 1128~\AA\ with \siiv. Figure~7 of Leitherer et al.
(2002) shows the synthetic spectra from 0 to 100~Myr obtained with a stellar
library of observed O stars in Starburst99 in the spectral range from 1200
to 1700~\AA: the metallicity effects can be directly assessed from this plot,
showing both $1\over 4$ and solar-metallicity models. For example, the emission
component of \civ\,$\lambda$1450 is clearly stronger in the
subsolar-metallicity model for
certain ages around 4~Myr. P$^{3+}$ has nearly the same ionization potential
(51~eV) as C$^{2+}$ (48~eV), so it is not improbable for these two lines to
behave similarly. Indeed, at age 3-5~Myr, the \pv\ line at 1118\AA\ shows the
strongest emission component at subsolar metallicity.

For the subsolar-metallicity models, it is easier to identify the
\siv\,$\lambda\lambda$1063,~1073,~1074 multiplet because the molecular hydrogen
contribution from the Magellanic Cloud has been removed in the LMC+SMC library.
In these models, a clear P~Cygni wind profile is seen around 3~Myr, i.e. when
supergiant stars are important contributors to the far-UV flux. The emission
peak of the line at 1063~\AA\ is partially masked by an interstellar
absorption from Ar\,{\sc i}\,$\lambda$1066.7. In the solar-metallicity
models, a P~Cygni wind profile for the \siv\ line at 1073 -- 1074~\AA\
is also seen around 3~Myr while the H$_2$ lines complicates the
interpretation of the component at 1063~\AA.  The effect of the metallicity
on the \ovi\,$\lambda\lambda$1032,~1038 doublet is more difficult to describe
because of the increasing importance of H$_2$ lines at shorter wavelength
in the solar-metallicity models and the proximity of Ly$\beta$.

\subsection{Continuous star formation} 
 
Synthetic spectra from 0 to 500~Myr for a stellar population generated in a
continuous star formation regime with a standard IMF at solar and subsolar
metallicity are shown in Figures~\ref{cont.sol.Salp.100} and
\ref{cont.sub.Salp.100}, respectively.
In these cases, an equilibrium between the stellar death and birth
is reached around 10~Myr for O stars and around 100~Myr for B stars
(see Fig.~\ref{nbstar}).
For the solar-metallicity models, at 10~Myr, the effects related to the
dilution problem are not observed because stars up to the spectral type B3
are included in the solar-metallicity library.
After 10~Myr, the solar-metallicity models are
all identical.  This clearly indicates that dilution effects by the
missing spectral groups in the library are negligible and it confirms that
the luminosity in the integrated
far-UV spectrum is dominated by the constant creation of massive OB stars.
At subsolar metallicity (Fig.~\ref{cont.sub.Salp.100}), an equilibrium for the
O stars is also reached around 10~Myr. However, the lack of
B-star spectra in the LMC+SMC library implies that dilution effects are
already present at relatively young ages. At 10~Myr, the subsolar-metallicity
model does not present an appreciable dilution effect and can still be
compared to the
observed galaxy spectra. For ages greater than 10-15~Myr, the dilution effect
is seen and the strength of the lines is not useful anymore to date the stellar
populations. This implies that in a continuous star-formation regime only
solar-metallicity
models are currently available to constrain the age of stellar populations
older than 10~Myr.

Figure~\ref{cont.salp.100} allows a direct comparison between the solar- and
subsolar-metallicity models for a standard continuous star formation at ages
of 0, 2, 5, 7, and 10~Myr. This plot can be used to assess the metallicity
effects in the dating of stellar populations with continuous star formation
for ages less than 10~Myr. Furthermore, the trends found for young stellar
populations can be extrapolated to older ages, when only solar-metallicity
synthetic spectra are available. The main striking differences between the
two spectral models are the decreased photospheric
(and interstellar) line blanketing at lower metallicity
(e.g. \siiv\,$\lambda\lambda$1122,~1128) and the weaker absorption components
of the wind profile of, for example, \pv\,$\lambda\lambda$1118,~1128 and
\ciii\,$\lambda$1176. It is interesting to notice how the very weak emission
component of these wind profiles are nearly the same for the two metallicities
and can thus be used with a certain confidence to extrapolate the
solar-metallicity
models at lower metallicity for ages greater than 10~Myr.

\subsection{Instantaneous vs. continuous} 
 
Models with an instantaneous and a continuous star formation at ages less
than a few Myr are almost indistinguishable (refer to
Fig.~\ref{inst.sol.Salp.100} and \ref{cont.sol.Salp.100} at solar metallicity,
or Fig.~\ref{inst.sub.Salp.100} and \ref{cont.sub.Salp.100} at subsolar
metallicity).
The difference starts to be clear around 3~Myr when, in the case of the
continuous star-forming rate, the flux from continuously-added new
massive stars
weakens the wind profiles of the first appearing O supergiants.
The strongest P~Cygni profiles are only seen for an instantaneous burst.
The presence of the \civ/\niv\,$\lambda$1169 (see \S3) near
\ciii\,$\lambda$1176 is also very specific to the type of stellar formation.
The photospheric feature of \civ/\niv\,$\lambda$1169 is
stronger for hot O stars and disappears underneath the broad and strong P~Cygni
\ciii\,$\lambda$1176 wind profile of O supergiants.
Therefore, only in the case of a continuous model do we keep observing this
\civ/\niv\ feature at all ages.

\subsection{IMF variations} 

In the first few Myr of the evolution of a star-forming region, the shape
of the stellar-wind lines has a strong dependence not only on the age, but
also on the IMF. Figure~\ref{IMF.alpha.inst.sol} shows how the synthesis
models for an instantaneous burst at solar metallicity change with the
slope of the IMF. For this comparison we have selected a burst of age 3~Myr,
when the influence of the IMF slope is particularly obvious. The wind lines of
\ciii\,$\lambda$1176 and \pv\,$\lambda\lambda$1118,~1128,
in the spectral window at $\lambda > 1110$~\AA\ where there are
practically no H$_2$ absorption lines, clearly show a stronger
emission component for a flatter IMF (i.e., $\alpha = 1.5$). This effect is a
direct consequence of a relative increase in the number of massive O stars with
their wind signatures with respect to less massive B stars with their stronger
photospheric absorption.  The same trend with the IMF slope is seen
for the other lines of \ovi\,$\lambda\lambda$1032,~1038 and
\siv\,$\lambda\lambda$1063,~1073,~1074, although not so clearly as these
fall in the wavelength spectral region with stronger contamination by
molecular hydrogen. The photospheric absorption feature of
\civ/\niv\,$\lambda$1169 is also enhanced for a flatter IMF slope
because hot O stars with stronger \civ/\niv\ absorption are relatively more
numerous.  In the case of a steeper IMF slope
(i.e., $\alpha = 3.3$), the opposite trend is observed.
For example, in Figure~\ref{IMF.alpha.inst.sol}, the photospheric
lines of \ciii\,$\lambda$1176 and \pv\,+\siiv\ at 1128~\AA\ get deeper
while the blueshifted wind absorption weakens.

Figure~\ref{IMF.Mup.inst.sol} shows models for an instantaneous burst at
2~Myr for a Salpeter IMF slope with upper mass limits of 30, 50, and
100~M$_\odot$. If no stars more massive than 30 -- 50~M$_{\odot}$ are present
in the stellar population, the P~Cygni profile for the wind lines disappears.
For the same reason, the
\civ/\niv\,$\lambda$1169 photospheric O star line gets weaker.
The photospheric contribution of \ciii\,$\lambda$1176 and \pv\,+\siiv\ at
1128~\AA\  is also enhanced by the higher percentage of lower mass stars.

The same trends with the IMF slope and upper mass limit can be seen at subsolar
metallicity, even if the lower metallicity renders the contrast much less
extreme. Similar arguments can be applied as well to the continuous
star-formation regime.

\subsection{Luminosities} 
 
All the spectra discussed in this paper are normalized to the unity level, but
were calculated in luminosity units as well. Both the normalized and the
absolute luminosities can be calculated and retrieved from the Starburst99
web site at www.stsci.edu/science/starburst99/.
The continuum always has the correct absolute flux level
since the evolutionary tracks and model
atmospheres with the appropriate
chemical composition are used for the synthetic spectrum.
Tables~5 and 6 report the evolution of the
monochromatic luminosity at 1160~\AA~ with age in our models with different
IMFs at solar and subsolar metallicity for an instantaneous and continuous
star-formation regime, respectively. This particular wavelength has been chosen
as representative of the continuum absolute flux, being almost free of strong
absorption lines. The values in these tables can be used to scale the
normalized spectra in the models presented in Figures~\ref{inst.sol.Salp.100}
to \ref{IMF.Mup.inst.sol} to an absolute flux scale.

It is important to stress that the far-UV continuum shape varies with age
and can be used as an additional age indicator for instantaneous bursts.
The reason is the following: the intrinsic stellar spectra below 1200~\AA~
are no longer close to the Rayleigh-Jeans regime, and age effects are no longer
negligible if most of the light comes from an instantaneous population.
Alternatively, in the case of continuous star formation the same spectral
region becomes much less age sensitive to population variations because an
equilibrium between forming and dying stars is reached quite
early in time (Leitherer et al. 2002). Figure~\ref{slopes} shows the
slope $\beta_{FUV}$ (from $F_\lambda~\alpha~\lambda^{\beta_{FUV}}$) calculated
in the continuum between 1044 and 1182~\AA.  The calculation of $\beta_{FUV}$
was done using synthetic spectra based on atmosphere models only for all
stars. If library spectra were to be taken into account, stellar features would
influence the definition of the continuum level in a none consistent way
at this point since cooler stars are not included in our libraries yet.
A strongly negative (blue) spectral slope indicates a young age.
This slope turns over as the population gets older and is sensitive to
the type of burst (instantaneous vs. continuous) and to the metallicity.
When comparing these slopes with a value obtained for the observed
spectra of a real stellar population, special care must be taken to
account for the absorption-line crowding, especially shortward of
$\sim$1110~\AA\ primarily due to the presence of the molecular hydrogen
and the Lyman-series lines.

\section{Conclusions} 
 
We have created stellar spectral libraries of hot stars in the
wavelength range 1003~--1183~\AA\ from \fuse\ archival data. These
new libraries complement our earlier work in the satellite-UV
longward of L$\alpha$, providing almost continuous wavelength
coverage from 1000 to 1800~\AA. In comparison with the longer
wavelength {\it HST} data, the \fuse\ spectra have 10 times higher
resolution and at the same time about the same S/N. The
library stars are located in the Milky Way, the LMC, and the
SMC and extend over a metallicity range of almost a factor of
10.

The spectral region below 1200~\AA\ shows strong line-blanketing
due to stellar-wind, stellar photospheric, and interstellar lines.
The stellar features generally originate from higher ionization
stages than the features found above 1200~\AA. The most prominent
transition is the \ovi\ resonance doublet at 1032, 1038~\AA\ which
displays a spectacular P~Cygni profile over a broad range of
spectral types. At the resolution afforded by \fuse, the blueshifted
absorption component of the P~Cygni profile is resolved from nearby
L$\beta$ and can be distinguished from the narrow interstellar
\cii\ at 1036~\AA. The (redshifted) emission component of its P~Cygni
profile is relatively unaffected by interstellar lines and provides
additional diagnostic power.
The \ciii\ $\lambda$1176 line is at the long-wavelength end of
the covered spectral range and can also be observed with
spectrographs optimized for wavelengths longward of L$\alpha$.
Surprisingly, the line has received relatively little attention in
the earlier literature. We find it a very good diagnostic of
the properties of hot stars. \ciii\ is not a resonance transition,
and consequently does not suffer from contamination by an
interstellar component. \ciii, like most other stellar lines,
has a pronounced metallicity dependence, either directly via opacity
variations, or indirectly via metallicity dependent stellar-wind
properties.

The \fuse\ wavelength range is particularly rich in interstellar lines
from molecular, atomic, and ionic transitions. Even at the spectral
resolution of \fuse, the blending of stellar and interstellar features
can be complex, and care is required when interpreting the spectra.
The numerous transitions of molecular hydrogen are dominant in
the Galactic stars, but less so in the LMC and SMC stars. The lower
column densities associated with the Clouds have allowed us to
model and remove the H$_2$ lines in a subset of the LMC/SMC stars and
to generate library stars virtually uncontaminated by H$_2$.

The libraries for different metallicities were integrated into the
LavalSB and Starburst99 synthesis codes. A suite of standard
synthetic spectra was generated to study the basic properties
of stellar population spectrum as a function of the most relevant
parameters. These model sets serve as a baseline for comparison
with young galaxy spectra, both observed locally in the far-UV or in the
distant universe when redshifted into the visual wavelength range.
Such a comparison will provide insight into the properties of
the stellar content and of the opacity of the intervening intergalactic
medium. Additionally, fully theoretical far-UV line spectra for
stars in any position on the Hertzsprung-Russell diagram will soon
become widely available. Our empirical set of spectra can provide
tests and guidelines for such theoretical approaches.

The readers are encouraged to explore a broader parameter range than
discussed in this paper by visiting www.stsci.edu/science/starburst99/
and running a set of tailored models.

\acknowledgments

CR acknowledges financial support from the Natural Sciences
and Engineering Research Council of Canada and the Université Laval.

\clearpage 
 
\noindent {\bf Captions:} 
 
\figcaption{Lines from molecular hydrogen in far-UV stellar spectra.
H$_2$ lines, along with other important interstellar and stellar features,
are identified in the Galaxy restframe.
The Magellanic Cloud spectra have not been deredshifted.
For the Magellanic stars, the spectrum corrected for the H$_2$ contribution
from the Cloud is shown (thick full lines) superposed to the
uncorrected data (dashed lines).
\label{fig0}}

\figcaption{Stellar indicators from the far-UV libraries
for groups of dwarf stars at various metallicities.
The stellar multiplets of \ovi\,$\lambda\lambda$1032,~1038,
\siv\,$\lambda\lambda$1163,~1173, 1174,
\pv\,$\lambda\lambda$1118,~1128, and \ciii\,$\lambda$1176
are presented for the groups O3, O5, O7, O9, and B3 ({\it top to bottom}).
Less important stellar lines of \siiv\,$\lambda\lambda$1123,~1128 and
\civ/\niv\,$\lambda$1169 are also indicated.
Interstellar features and molecular hydrogen lines are identified.
Spectra were taken from the solar-metallicity library ({\it dark solid lines}),
the LMC+SMC library ({\it thin solid lines}), the LMC library
({\it dotted lines}), and the SMC library ({\it dashed lines}).
The narrow spike at 1168.7\,\AA\ often seen in the LMC and SMC
libraries is a scattered \hei\ solar emission.
\label{fig1}}

\figcaption{Same as Figure~2 but for giant stars.
\label{fig2}}

\figcaption{Same as Figure~2 but for supergiant stars.
\label{fig3}}

\figcaption{Number of hot stars as a function of the stellar population age.
Instantaneous bursts ({\it left}) and continuous bursts ({\it right}) are
presented for two metallicities: solar ({\it top}) and subsolar ({\it bottom}).
Model parameters are: $\alpha = 2.35$, $M_{low}$~=~1~M$_{\odot}$,
and $M_{upp}$~=~100~M$_{\odot}$.
For the instantaneous burst, $10^6$~M$_\odot$ were converted into stars
at the initial time of the star formation. For the continuous burst, a rate
of star formation equal to 1~M$_\odot$~yr$^{-1}$ was adopted.
O stars ({\it dark solid lines}) and B stars ({\it thin solid lines}) have
been grouped according to their luminosity class. Numbers of WR stars
({\it dotted lines}) are also shown.
\label{nbstar}}

\figcaption{Evolution of a synthetic far-UV spectrum between 0 and 10~Myr.
Model parameters are: instantaneous burst, Z$_{\odot}$, $\alpha$~=~2.35,
$M_{low}$~=~1~M$_{\odot}$, and $M_{upp}$~=~100~M$_{\odot}$.
The spectra have been normalized to a continuum equal to unity to
facilitate the line comparison.
Stellar features ({\it bottom}), H$_2$ lines ({\it middle}),
and other interstellar absorptions ({\it top, dotted lines}) are indicated.
The emission line in the spectra of the Galactic stars around 1026~\AA\ is
geocoronal.
\label{inst.sol.Salp.100}}

\figcaption{Same as Fig.~\ref{inst.sol.Salp.100} but for subsolar
metallicity ($1\over 4$~Z$_{\odot}$). Some weak emission around 1025~\AA\
is due to  geocoronal L$\beta$ in the LMC/SMC library stars (the two galaxies
have slightly different heliocentric and systemic velocities with respect to
each other).
\label{inst.sub.Salp.100}}

\figcaption{Evolution of a synthetic far-UV spectrum between 0 and 500~Myr.
Model parameters are: continuous star formation, Z$_{\odot}$, $\alpha$~=~2.35,
$M_{low}$~=~1~M$_{\odot}$, and $M_{upp}$~=~100~M$_{\odot}$.
The spectra have been normalized to a continuum equal to unity to
facilitate the line comparison.
Stellar features ({\it bottom}), H$_2$ lines ({\it middle}), and
other interstellar absorptions ({\it top, dotted lines}) are indicated.
\label{cont.sol.Salp.100}}

\figcaption{Same as Fig.~\ref{cont.sol.Salp.100} but for subsolar metallicity
($1\over 4$~Z$_{\odot}$).
\label{cont.sub.Salp.100}}

\figcaption{Comparison of synthetic spectra between 0 and 10~Myr for continuous
star formation.
Model parameters are:
$\alpha$~=~2.35, $M_{low}$~=~1~M$_{\odot}$, $M_{upp}$~=~100~M$_{\odot}$,
and Z~=~$1\over 4$~Z$_{\odot}$ ({\it solid lines}), or
Z~=~Z$_{\odot}$ ({\it dotted lines}).
The spectra have been normalized to a continuum equal to unity to
facilitate the line comparison.
Stellar features ({\it bottom}), H$_2$ lines ({\it middle}), and other
interstellar absorptions ({\it top, dotted lines}) are identified.
Only the spectral region for $\lambda~>~1110$~\AA\ is shown.
\label{cont.salp.100}}

\figcaption{3~Myr instantaneous burst at solar metallicity for different
slopes of the IMF. Model parameters are: $M_{low}$~=~1~M$_{\odot}$,
$M_{upp}$~=~100~M$_{\odot}$, and $\alpha$~=~2.35 ({\it solid line}),
$\alpha$~=~3.3 ({\it dotted line}), or $\alpha$~=~1.5 ({\it dashed line}).
The spectra have been normalized to a continuum equal to unity to
facilitate the line comparison.
Stellar features are identified at the {\it bottom}.
\label{IMF.alpha.inst.sol}}

\figcaption{2~Myr instantaneous burst at solar metallicity for different
upper cut-off masses. Model parameters are: $\alpha~=~2.35$,
$M_{low}$~=~1~M$_{\odot}$, and $M_{upp}$~=~100~M$_{\odot}$ ({\it solid line}),
$M_{upp}$~=~50~M$_{\odot}$ ({\it dotted line}), or
$M_{upp}$~=~30~M$_{\odot}$ ({\it dashed line}).
The spectra have been normalized to a continuum equal to unity to
facilitate the line comparison.
Stellar features are identified at the {\it bottom}.
\label{IMF.Mup.inst.sol}}

\figcaption{$\beta_{FUV}$, the slope of the far-UV spectrum between 1044
and 1182~\AA. Values for four models are plotted as a function of time:
instantaneous ({\it solid lines}) and continuous burst ({\it dashed
lines}) with solar and subsolar metallicity. Model parameters are:
$\alpha~=~2.35$, $M_{low}$~=~1~M$_{\odot}$, and $M_{upp}$~=~100~M$_{\odot}$.
\label{slopes}}

\clearpage

\begin{deluxetable}{lcclcrcc}
\renewcommand{\arraystretch}{0.95}
\tabletypesize{\scriptsize}
\tablecolumns{8}
\tablewidth{0pc}
\tablecaption{Fundamental Parameters of Galactic Stars}
\tablehead{ \colhead{Star} & \colhead{R.A. (J2000)} & \colhead{Dec (J2000)}
& \colhead{Spectral Type} & \colhead{Ref} & \colhead{V} & \colhead{E(B$-$V)}
& \colhead{ID} \\
& \colhead{(h m s)} & \colhead{($^\circ$ $'$ $''$)} &  &  &  &  &  }
\startdata
HD 3827     & 00 41 12 &   +39 36 13 & B0.7V((n))    & 01 & 8.01 & 0.02 & P1010302  \\
HD 4004     & 00 43 28 &   +64 45 35 & WN4b          & 02 & 10.54 & 0.76 & P1170301 \\
HD 5005A    & 00 52 49 &   +56 37 39 & O6.5V((f))    & 03 & 7.76 & 0.41 & P1020102  \\
HD 12323    & 02 02 30 &   +55 37 26 & ON9V          & 04 & 8.90 & 0.29 & P1020202  \\
HD 12740    & 02 06 11 &   +49 09 22 & B1.5II        & 01 & 7.94 & 0.18 & P1010701  \\
HD 13268    & 02 11 29 &   +56 09 31 & O8V           & 05 & 8.18 & 0.44 & P1020304  \\
HD 13745    & 02 15 45 &   +55 59 46 & O9.7II((n))   & 04 & 7.83 & 0.45 & P1020404  \\
HD 15137    & 02 27 59 &   +52 32 57 & O9.5II-III(n) & 04 & 7.86 & 0.35 & P1020602  \\
HD 15642    & 02 32 56 &   +55 19 39 & O9.5III:n     & 01 & 8.51 & 0.37 & P1020702  \\
HD 22586    & 03 35 37 & $-$52 33 23 & B2III         & 06 & 8.03 & 0.05 & P1011101  \\
HD 27778    & 04 23 59 &   +24 18 03 & B3V           & 07 & 6.36 & 0.42 & P1160301  \\
HD 30677    & 04 50 03 &   +08 24 28 & B1II-III((n)) & 01 & 6.84 & 0.23 & P1020801  \\
HD 39680    & 05 54 44 &   +13 51 17 & O6V:[n]pe var & 08 & 7.99 & 0.34 & P1020901  \\
HD 42088    & 06 09 39 &   +20 29 15 & O6.5V         & 03 & 7.54 & 0.38 & P1021101  \\
HD 42401    & 06 10 59 &   +11 59 40 & B2V           & 09 & 7.35 & 0.18 & P1021201  \\
HD 45314    & 06 27 15 &   +14 53 22 & O9:pe         & 10 & 6.64 & 0.46 & P1021301  \\
HD 46150    & 06 31 55 &   +04 56 34 & O5V((f))      & 11 & 6.74 & 0.45 & P1021401  \\
HD 47360    & 06 38 23 &   +04 37 27 & B0.5V         & 10 & 8.19 & 0.41 & P1021504  \\
HD 47417    & 06 38 47 &   +06 54 06 & B0IV          & 10 & 6.97 & 0.31 & P1021601  \\
HD 60369    & 07 33 01 & $-$28 19 32 & O9IV          & 12 & 8.15 & 0.30 & P1050201  \\
HD 61347    & 07 38 16 & $-$13 51 02 & O9Ib          & 10 & 8.43 & 0.45 & P1022001  \\
HD 63005    & 07 45 49 & $-$26 29 31 & O6V           & 12 & 9.13 & 0.27 & P1022101  \\
HD 66788    & 08 04 08 & $-$27 29 09 & O8V           & 13 & 9.45 & 0.22 & P1011801  \\
HD 72088    & 08 29 12 & $-$44 53 05 & B3III-IV      & 14 & 9.07 & 0.23 & A1290601  \\
HD 74194    & 08 40 48 & $-$45 03 31 & O8.5Ib(f)     & 04 & 7.57 & 0.50 & P1022404  \\
HD 74662    & 08 43 18 & $-$48 20 43 & B3V           & 14 & 8.82 & 0.13 & A1290201  \\
HD 74711    & 08 43 47 & $-$46 47 56 & B1III         & 12 & 7.11 & 0.33 & P1022501  \\
HD 74920    & 08 45 10 & $-$46 02 19 & O7IIIn        & 15 & 7.53 & 0.34 & P1022601  \\
HD 75309    & 08 47 27 & $-$46 27 04 & B1IIp         & 12 & 7.86 & 0.25 & P1022701  \\
HD 88115    & 10 07 31 & $-$62 39 12 & B1.5IIn       & 12 & 8.30 & 0.16 & P1012301  \\
HD 89137    & 10 15 40 & $-$51 15 25 & O9.5III(n)p   & 16 & 7.98 & 0.23 & P1022801  \\
HD 90087    & 10 22 20 & $-$59 45 20 & O9IIn         & 12 & 7.76 & 0.28 & P1022901  \\
HD 91597    & 10 33 01 & $-$60 50 41 & B1IIIne       & 12 & 9.84 & 0.30 & P1023002  \\
HD 91651    & 10 33 30 & $-$60 07 35 & O9V:n         & 04 & 8.84 & 0.29 & P1023102  \\
HD 91824    & 10 34 46 & $-$58 09 22 & O7V((f))      & 11 & 8.16 & 0.26 & A1180802  \\
HD 92702    & 10 41 00 & $-$57 36 02 & B1Iab         & 17 & 8.14 & 0.38 & S5130301  \\
HD 92809    & 10 41 38 & $-$58 46 19 & WC6           & 18 & 9.08 & 0.22 & P1170401  \\
HD 93028    & 10 43 15 & $-$60 12 04 & O9V           & 17 & 8.39 & 0.24 & A1180902  \\
HD 93129A   & 10 43 57 & $-$59 32 51 & O2If*         & 19 & 8.84 & 0.58 & P1170202  \\
HD 93146    & 10 43 59 & $-$60 05 11 & O6.5V((f))    & 04 & 8.45 & 0.35 & P1023301  \\
HD 93204    & 10 44 32 & $-$59 44 30 & O5V((f))      & 20 & 8.48 & 0.41 & P1023501  \\
HD 93205    & 10 44 33 & $-$59 44 15 & O3.5V((f+))   & 19 & 7.76 & 0.40 & P1023601  \\
HD 93206    & 10 44 23 & $-$59 59 36 & O9.7Ib:(n)    & 21 & 6.24 & 0.38 & P1023401  \\
HD 93222    & 10 44 36 & $-$60 05 29 & O7 III((f))   & 22 & 8.11 & 0.37 & P1023701  \\
HD 93250    & 10 44 45 & $-$59 33 54 & O3.5V((f))    & 19 & 7.37 & 0.48 & P1023801  \\
HD 93827    & 10 48 31 & $-$60 56 10 & B1Ibn         & 12 & 9.31 & 0.23 & P1023901  \\
HD 93840    & 10 49 08 & $-$46 46 42 & B2Ib          & 13 & 7.77 & 0.14 & P1012701  \\
HD 93843    & 10 48 37 & $-$60 13 25 & O5III(f)var   & 04 & 7.30 & 0.28 & P1024001  \\
HD 94493    & 10 53 15 & $-$60 48 52 & B1Ib          & 12 & 7.23 & 0.20 & P1024101  \\
HD 96548    & 11 06 17 & $-$65 30 34 & WN8           & 23 & 7.70 & 0.36 & P1170501  \\
HD 96715    & 11 07 32 & $-$59 57 48 & O4V((f))      & 24 & 8.27 & 0.42 & P1024301  \\
HD 96917    & 11 08 42 & $-$57 03 57 & O8.5Ib(f)     & 04 & 7.07 & 0.37 & P1024401  \\
HD 97471    & 11 12 07 & $-$58 48 14 & B0V           & 17 & 9.30 & 0.29 & A1180404  \\
HD 97913    & 11 14 54 & $-$59 10 29 & B0.5IVn       & 12 & 8.80 & 0.32 & P1221701  \\
HD 99857    & 11 28 27 & $-$66 29 21 & B0.5Ib        & 12 & 7.45 & 0.33 & P1024501  \\
HD 99890    & 11 29 05 & $-$56 38 38 & B0IIIn        & 12 & 8.28 & 0.24 & P1024601  \\
HD 100276   & 11 31 48 & $-$60 36 22 & B0.5Ib        & 12 & 7.16 & 0.26 & P1024801  \\
HD 101131   & 11 37 48 & $-$63 19 23 & O6V((f))      & 04 & 7.16 & 0.34 & P1024901  \\
HD 101190   & 11 38 10 & $-$63 11 49 & O6V((f))      & 04 & 7.27 & 0.36 & P1025001  \\
HD 100199   & 11 31 07 & $-$62 56 48 & B0IIIne       & 12 & 8.14 & 0.30 & P1221801  \\
HD 100213   & 11 31 10 & $-$65 44 32 & O8.5Vn        & 12 & 8.22 & 0.34 & P1024701  \\
HD 101298   & 11 39 03 & $-$63 25 46 & O6V((f))      & 04 & 8.05 & 0.38 & P1025201  \\
HD 101413   & 11 39 45 & $-$63 28 39 & O8V           & 04 & 8.35 & 0.36 & P1025301  \\
HD 102552   & 11 47 56 & $-$60 33 54 & B1IIIn        & 12 & 8.69 & 0.30 & P1025501  \\
HD 103779   & 11 56 57 & $-$63 14 56 & B0.5Iab       & 12 & 7.20 & 0.21 & P1025601  \\
HD 104705   & 12 03 24 & $-$62 41 45 & B0Ib          & 12 & 7.76 & 0.26 & P1025701  \\
HD 104994   & 12 05 18 & $-$62 03 09 & WN3p          & 23 & 10.93& 0.34 & S5160101  \\
HD 110432   & 12 42 50 & $-$63 03 31 & B2pe          & 25 & 5.24 & 0.36 & P1161401  \\
HD 114441   & 13 11 29 & $-$55 21 24 & B2Vne         & 26 & 8.02 & 0.36 & P1025801  \\
HD 114444   & 13 13 04 & $-$75 18 49 & B2III         & 06 & 10.32& 0.19 & A1180501  \\
HD 115071   & 13 16 04 & $-$62 35 00 & B0.5Vn        & 12 & 7.94 & 0.49 & P1025901  \\
HD 115473   & 13 18 28 & $-$58 08 14 & WC4           & 18 & 9.98 & 0.53 & P1170701  \\
HD 116538   & 13 25 11 & $-$51 50 29 & B2IVn         & 06 & 7.92 & 0.17 & P1026001  \\
HD 116781   & 13 27 25 & $-$62 38 55 & B0IIIne       & 12 & 7.60 & 0.43 & P1026101  \\
HD 116852   & 21 32 27 &   +10 08 19 & O9III         & 10 & 14.60& 0.07 & P1013801  \\
HD 118571   & 13 39 15 & $-$60 59 01 & B0.5IVn       & 12 & 8.76 & 0.26 & P1222001  \\
HD 118969   & 13 42 12 & $-$63 42 50 & B1.5V         & 26 & 10.00& 0.36 & P1222101  \\
HD 119608   & 13 44 31 & $-$17 56 13 & B1Ib          & 10 & 7.51 & 0.14 & P1014201  \\
HD 121800   & 13 55 15 &   +66 07 00 & B1.5V         & 27 & 9.11 & 0.08 & P1014401  \\
HD 121968   & 13 58 51 & $-$02 54 53 & B1V           & 28 & 10.25& 0.09 & P1014501  \\
HD 124979   & 14 18 11 & $-$51 30 13 & O8.5III       & 29 & 8.53 & 0.40 & P1026301  \\
HD 125924   & 14 22 43 & $-$08 14 53 & B2IV          & 06 & 9.68 & 0.05 & P1014701  \\
HD 134411   & 15 11 08 & $-$39 51 50 & B2Vn          & 06 & 9.56 & 0.06 & P1026501  \\
HD 146813   & 16 15 14 &   +55 47 58 & B1.5          & 30 & 9.06 & 0.02 & P1014901  \\
HD 151805   & 16 51 35 & $-$41 46 35 & B1Ib          & 17 & 8.86 & 0.32 & P1026602  \\
HD 151932   & 16 52 19 & $-$41 51 15 & WN7           & 23 & 6.49 & 0.50 & P1170801  \\
HD 152218   & 16 53 59 & $-$41 42 52 & O9.5IV(n)     & 04 & 7.61 & 0.47 & P1015402  \\
HD 152233   & 16 54 03 & $-$41 47 29 & O6III:(f)p    & 11 & 6.59 & 0.45 & P1026702  \\
HD 152248   & 16 54 09 & $-$41 49 30 & O7Ib:(n)(f)p  & 11 & 6.10 & 0.46 & P1026801  \\
HD 152623   & 16 56 14 & $-$40 39 36 & O7V(n)((f))   & 11 & 6.67 & 0.40 & P1027001  \\
HD 152723   & 16 56 54 & $-$40 30 43 & O6.5IIIf      & 11 & 7.16 & 0.47 & P1027102  \\
HD 153426   & 17 01 12 & $-$38 12 12 & O9II-III      & 11 & 7.47 & 0.45 & P1027202  \\
HD 154368   & 17 06 28 & $-$35 27 04 & O9.5Iab       & 04 & 6.13 & 0.55 & P1161901  \\
HD 156292   & 17 18 45 & $-$42 53 30 & O9.5III       & 04 & 7.49 & 0.56 & P1027402  \\
HD 156385   & 17 19 29 & $-$45 38 24 & WC7.5         & 31 & 7.45 & 0.08 & P1170901  \\
HD 157857   & 17 26 17 & $-$10 59 34 & O6.5III(f)    & 11 & 7.78 & 0.49 & P1027501  \\
HD 158243   & 17 31 07 & $-$53 28 42 & B1I(ab)       & 06 & 8.15 & 0.19 & P1015601  \\
HD 158661   & 17 31 12 & $-$17 08 31 & B0.5Ib        & 10 & 8.20 & 0.42 & P1222201  \\
HD 160993   & 17 45 17 & $-$45 38 13 & B1Iab         & 13 & 7.71 & 0.21 & P1015701  \\
HD 163522   & 17 58 35 & $-$42 29 09 & B1Ia          & 12 & 8.46 & 0.19 & P1015801  \\
HD 163758   & 17 59 28 & $-$36 01 15 & O6.5Iaf       & 04 & 7.32 & 0.34 & P1015901  \\
HD 164270   & 18 01 43 & $-$32 42 55 & WC9           & 18 & 9.01 & 0.17 & P1171001  \\
HD 164816   & 18 03 56 & $-$24 18 45 & O9.5III-IV(n) & 04 & 7.08 & 0.31 & P1016001  \\
HD 164906   & 18 04 25 & $-$24 23 09 & B1IVpe        & 26 & 7.42 & 0.42 & P1027701  \\
HD 165052   & 18 05 10 & $-$24 23 54 & O6.5V(n)((f)) & 04 & 6.86 & 0.42 & P1027801  \\
HD 165246   & 18 06 04 & $-$24 11 44 & O8V(n)        & 04 & 7.71 & 0.40 & P1050301  \\
HD 165763   & 18 08 28 & $-$21 15 11 & WC5           & 18 & 8.25 & 0.13 & P1171101  \\
HD 165955   & 18 09 57 & $-$34 52 06 & B1Vnp         & 12 & 9.18 & 0.15 & P1027901  \\
HD 166546   & 18 11 57 & $-$20 25 24 & O9.5II-III    & 04 & 7.24 & 0.34 & P1222501  \\
HD 167402   & 18 16 18 & $-$30 07 29 & B0Ib          & 12 & 8.95 & 0.23 & P1016201  \\
HD 167771   & 18 17 28 & $-$18 27 48 & O7III(n)((f)) & 11 & 6.54 & 0.44 & P1028101  \\
HD 167971   & 18 18 05 & $-$12 14 32 & O8Ib(f)p      & 11 & 7.31 & 1.08 & P1162101  \\
HD 168076   & 18 18 36 & $-$13 48 02 & O4V((f))      & 04 & 8.21 & 0.55 & P1162201  \\
HD 168080   & 18 18 46 & $-$18 10 19 & B0.5II        & 10 & 7.61 & 0.38 & P1222701  \\
HD 168941   & 18 23 25 & $-$26 57 10 & O9.5II-III    & 32 & 9.34 & 0.37 & P1016501  \\
HD 169673   & 18 26 23 & $-$15 37 48 & B1II          & 10 & 7.34 & 0.31 & P1050501  \\
HD 172140   & 18 39 48 & $-$29 20 21 & B0.5III       & 06 & 9.93 & 0.25 & P1016602  \\
HD 173502   & 18 46 55 & $-$29 57 34 & B1II          & 12 & 9.68 & 0.19 & P1016701  \\
HD 175754   & 18 57 36 & $-$19 09 11 & O8II((f))     & 11 & 7.01 & 0.23 & P1016802  \\
HD 175876   & 18 58 10 & $-$20 25 25 & O6.5III(n)(f) & 04 & 6.95 & 0.22 & P1016902  \\
HD 178487   & 19 09 14 & $-$10 13 03 & B0Ib          & 13 & 8.66 & 0.40 & P1017201  \\
HD 177989   & 19 07 37 & $-$18 43 34 & B0III         & 33 & 9.33 & 0.25 & P1017101  \\
HD 179406   & 19 12 40 & $-$07 56 22 & B3V           & 34 & 5.36 & 0.24 & P2160701  \\
HD 179407   & 19 12 52 & $-$12 34 57 & B0.5Ib        & 33 & 9.41 & 0.31 & P1017301  \\
HD 183899   & 19 32 45 & $-$26 09 46 & B2III         & 06 & 9.80 & 0.16 & P1017601  \\
HD 187282   & 19 48 32 &   +18 12 04 & WN4           & 23 & 10.56& 0.18 & P1171201  \\
HD 187459   & 19 48 50 &   +33 26 14 & B0.5III       & 34 & 6.44 & 0.42 & P1028201  \\
HD 190429   & 20 03 29 &   +36 01 29 & O4If+         & 22 & 6.56 & 0.51 & P1028401  \\
HD 191495   & 20 08 53 &   +35 30 46 & B0IV-V(n)     & 01 & 8.26 & 0.40 & P1222901  \\
HD 191765   & 20 10 14 &   +36 10 36 & WN6           & 23 & 8.31 & 0.45 & P1171301  \\
HD 191877   & 20 11 21 &   +21 52 31 & B1Ib          & 34 & 6.28 & 0.18 & P1028701  \\
HD 192035   & 20 10 49 &   +47 48 47 & B0III-IV(n)   & 01 & 8.20 & 0.36 & P1028601  \\
HD 192103   & 20 11 53 &   +36 11 50 & WC8           & 18 & 8.09 & 0.53 & P1171401  \\
HD 192639   & 20 14 30 &   +37 21 13 & O7Ib(f)       & 11 & 7.11 & 0.64 & P1162401  \\
HD 193077   & 20 16 60 &   +37 25 24 & WN5           & 23 & 7.97 & 0.66 & P1171501  \\
HD 195455   & 20 32 14 & $-$24 04 03 & B0.5III       & 06 & 9.20 & 0.10 & P1017801  \\
HD 195965   & 20 32 25 &   +48 12 59 & B0V           & 35 & 6.98 & 0.25 & P1028803  \\
HD 199579   & 20 56 34 &   +44 55 29 & O6V((f))      & 04 & 6.01 & 0.55 & P1162501  \\
HD 201345   & 21 07 55 &   +33 23 49 & ON9V          & 04 & 7.66 & 0.18 & P1223001  \\
HD 201638   & 21 09 53 &   +35 29 30 & B0.5Ib        & 36 & 9.10 & 0.08 & P1018001  \\
HD 203938   & 21 23 50 &   +47 09 52 & B0.5IV        & 10 & 7.45 & 0.74 & P1162601  \\
HD 210121   & 22 08 11 & $-$03 31 52 & B3V           & 37 & 7.67 & 0.55 & P1163001  \\
HD 210809   & 22 11 38 &   +52 25 47 & O9Iab         & 04 & 7.54 & 0.33 & P1223102  \\
HD 210839   & 22 11 30 &   +59 24 52 & O6I(n)f       & 11 & 5.06 & 0.62 & P1163101  \\
HD 212044   & 22 20 22 &   +51 51 39 & B1:V:pnne     & 10 & 6.98 & 0.30 & P1223401  \\
HD 215733   & 22 47 02 &   +17 13 59 & B1II          & 10 & 7.34 & 0.11 & P1018601  \\
HD 216044   & 22 48 43 &   +55 07 33 & B0III-IV      & 01 & 8.51 & 0.37 & P1223801  \\
HD 218915   & 23 11 06 &   +53 03 30 & O9.5Iab       & 11 & 7.20 & 0.29 & P1018801  \\
HD 224151   & 23 55 33 &   +57 24 43 & B0.5II-III    & 34 & 6.00 & 0.48 & P1224101  \\
HD 224257   & 23 56 25 &   +55 59 25 & B0.2IV        & 01 & 7.98 & 0.23 & P1050601  \\
HD 224868   & 00 01 21 &   +60 50 21 & B0Ib          & 38 & 7.25 & 0.37 & P1220201  \\
HD 225757   & 19 46 41 &   +34 39 14 & B1IIIn        & 39 & 10.59& 0.22 & P1017703  \\
HD 239683   & 21 29 53 &   +57 48 57 & B3IV          & 40 & 9.32 & 0.48 & A1181001  \\
HDE 232522  & 01 46 02 &   +55 19 54 & B1II          & 10 & 8.67 & 0.23 & P1220101  \\
HDE 233622  & 09 21 33 &   +50 05 56 & B2V           & 41 & 10.01& 0.03 & P1012102  \\
HDE 235783  & 22 17 07 &   +54 30 27 & B1Ib          & 10 & 8.68 & 0.36 & P1223301  \\
HDE 235874  & 22 32 59 &   +51 12 56 & B3III         & 39 & 9.64 & 0.20 & P1223701  \\
HDE 303308  & 10 45 06 & $-$59 40 05 & O4V((f+))     & 19 & 8.17 & 0.45 & P1221602  \\
HDE 308813  & 11 37 58 & $-$63 18 58 & O9.5V         & 42 & 9.28 & 0.34 & P1221901  \\
HDE 315021  & 18 04 35 & $-$24 19 51 & B2V           & 17 & 8.63 & 0.26 & P1222401  \\
HDE 332407  & 19 41 19 &   +29 08 40 & B1Ibp         & 35 & 8.50 & 0.48 & P1222801  \\
CPD -59$^{\circ}$2600& 10 44 41 & $-$59 46 55 & O6V((f))      & 04 & 8.61 & 0.53 & P1221401 \\
CPD -59$^{\circ}$2603& 10 44 47 & $-$59 43 51 & O7V((f))      & 04 & 8.77 & 0.46 & P1221501 \\
CPD -69$^{\circ}$1743& 13 00 33 & $-$70 12 35 & B1Vn          & 43 & 9.38 & 0.30 & P1013701 \\
CPD -72$^{\circ}$1184& 11 59 00 & $-$73 25 46 & B0III         & 33 & 10.68& 0.23 & S5140101 \\
CPD -74$^{\circ}$1569& 16 50 50 & $-$74 32 20 & O9.5V         & 33 & 10.15& 0.13 & P1015301 \\
BD +35$^{\circ}$4258 & 20 46 12 &   +35 32 26 & B0.5Vn        & 10 & 9.41 & 0.29 & P1017901 \\
BD +38$^{\circ}$2182 & 10 49 12 &   +38 00 14 & B3V           & 41 & 11.25& 0.00 & P1012801 \\
BD +53$^{\circ}$2820 & 22 13 49 &   +54 24 34 & B0IVn         & 35 & 9.95 & 0.40 & P1223201 \\
BD +56$^{\circ}$524  & 02 19 06 &   +57 07 33 & B1Vn          & 44 & 9.75 & 0.60 & A1181112 \\
LS 277      & 07 16 12 & $-$08 31 14 & B1V           & 45 & 9.78 & 0.27 & P1220801  \\
JL 212      & 00 49 01 & $-$56 05 48 & B2V           & 45 & 10.20& 0.13 & P1010401  \\
\enddata
\tablerefs{ 1.~Walborn (1971a); 2.~Smith et~al. (1996); 3.~Morrell et al. (1991);
4.~Walborn (1973a); 5.~Mathys (1989); 6.~Hill (1970); 7.~Morguleff \& Gerbaldi (1975);
8.~Walborn et al. (1985); 9.~Walborn \& Fitzpatrick (1990); 10.~Morgan et al. (1955);
11.~Walborn (1972); 12.~Garrison et~al. (1977); 13.~MacConnell \& Bidelman (1976);
14.~Houk (1978); 15.~Vijapurkar \& Drilling (1993); 16.~Walborn (1976);
17.~Reed \& Beatty (1995); 18.~Smith et~al. (1990); 19.~Walborn et~al. (2002b);
20.~Massey \& Johnson (1993); 21.~Walborn (1973b); 22.~Walborn (1971b);
23.~Smith et~al. (1996); 24.~Cruz-Gonzalez et~al. (1974); 25.~Th\'e et~al. (1986);
26.~Jaschek et~al. (1964); 27.~Dworetsky et~al. (1982); 28.~Little et~al. (1995);
29.~Howarth \& Prinja (1989); 30.~Danly (1989); 31.~Hiltner \& Schild (1966);
32.~Walborn (1982); 33.~Hill et~al. (1974); 34.~Rountree Lesh (1968);
35.~Hiltner (1956); 36.~Bidelman (1951); 37.~Welty \& Fowler (1992);
38.~Barbier \& Boulon (1960); 39.~Crampton et~al. (1976); 40.~Garrison \& Kormendy (1976);
41.~Ryans et~al. (1997); 42.~Schild (1970); 43.~Walborn \& Fitzpatrick (2000);
44.~Crawford et~al. (1970); 45.~PI of the FUSE proposal.}
\end{deluxetable}

\begin{deluxetable}{lcclcrcc}
\renewcommand{\arraystretch}{0.95}
\tabletypesize{\scriptsize}
\tablecolumns{8}
\tablewidth{0pc}
\tablecaption{Fundamental Parameters of Stars in the Magellanic Clouds}
\tablehead{ \colhead{Star} & \colhead{R.A. (J2000)} & \colhead{Dec (J2000)}
& \colhead{Spectral Type} & \colhead{Ref} & \colhead{V} & \colhead{E(B$-$V)}
& \colhead{ID} \\
& \colhead{(h m s)} & \colhead{($^\circ$ $'$ $''$)} &  &  &  &  &  }
\startdata
BI 170     &  05 26 47 & $-$69 06 11 & O9.5Ib      & 01 & 13.09 & 0.13 & P1173701  \\
BI 173     &  05 27 10 & $-$69 07 56 & O8III       & 02 & 13.00 & 0.17 & P1173202  \\
BI 208     &  05 33 57 & $-$67 24 20 & O7V         & 02 & 14.02 & 0.03 & P1172704  \\
BI 229     &  05 35 32 & $-$66 02 37 & O7III       & 02 & 12.95 & 0.15 & P1172801  \\
BI 272     &  05 44 23 & $-$67 14 29 & O7III       & 03 & 13.20 & 0.17 & P1172902  \\
HD 32109   &  04 55 31 & $-$67 30 00 & WN4b        & 04 & 13.87 & 0.00 & P1174402  \\
HD 33133   &  05 03 10 & $-$66 40 53 & WN8         & 04 & 12.69 & 0.08 & P1174501  \\
HD 37026   &  05 30 12 & $-$67 26 08 & WC4         & 05 & 14.30 & 0.08 & P1175001  \\
HD 269810  &  05 35 13 & $-$67 33 27 & O2III(f*)   & 06 & 12.26 & 0.14 & P1171603  \\
HDE 269582 &  05 27 52 & $-$68 59 08 & WN10        & 07 & 11.88 & 0.09 & P1174701  \\
HDE 269687 &  05 31 25 & $-$69 05 38 & WN11        & 07 & 11.90 & 0.10 & P1174801  \\
HDE 269927 &  05 38 58 & $-$69 29 19 & WN9         & 08 & 12.48 & 0.19 & P1174601  \\
MK 42      &  05 38 42 & $-$69 05 54 & O3If/WN     & 09 & 10.96 & 0.45 & P1171802  \\
SK -65$^{\circ}$21  &  05 01 22 & $-$65 41 48 & O9.7Iab     & 10 & 12.02 & 0.20 & P1030904  \\
SK -65$^{\circ}$22  &  05 01 23 & $-$65 52 33 & O6Iaf+      & 01 & 12.07 & 0.20 & P1031002  \\
SK -66$^{\circ}$18  &  04 55 59 & $-$65 58 30 & O6V((f))    & 11 & 13.50 & 0.12 & A0490102  \\
SK -66$^{\circ}$100 &  05 27 46 & $-$66 55 15 & O6II(f)     & 10 & 13.26 & 0.12 & P1172303  \\
SK -66$^{\circ}$169 &  05 36 54 & $-$66 38 24 & O9.7Ia+     & 08 & 11.56 & 0.16 & P1173801  \\
SK -66$^{\circ}$172 &  05 37 05 & $-$66 21 35 & O2III(f*)+OB& 06 & 13.13 & 0.21 & P1172201  \\
SK -66$^{\circ}$185 &  05 42 30 & $-$66 18 10 & B0Iab       & 02 & 13.11 & 0.05 & A0490902  \\
SK -67$^{\circ}$05  &  04 50 18 & $-$67 39 37 & O9.7Ib      & 08 & 11.34 & 0.15 & P1030704  \\
SK -67$^{\circ}$14  &  04 54 31 & $-$67 15 24 & B1.5Ia      & 08 & 11.52 & 0.10 & P1174202  \\
SK -67$^{\circ}$28  &  04 58 39 & $-$67 11 18 & B0.7Ia      & 08 & 12.28 & 0.10 & A0490202  \\
SK -67$^{\circ}$46  &  05 07 01 & $-$67 37 29 & B1.5I       & 12 & 12.34 & 0.14 & A0491501  \\
SK -67$^{\circ}$69  &  05 14 20 & $-$67 08 03 & O4III(f)    & 13 & 13.09 & 0.16 & P1171703  \\
SK -67$^{\circ}$76  &  05 20 05 & $-$67 21 08 & B0Ia        & 14 & 12.42 & 0.20 & P1031201  \\
SK -67$^{\circ}$101 &  05 25 56 & $-$67 30 28 & O8II((f))   & 15 & 12.63 & 0.14 & P1173403  \\
SK -67$^{\circ}$104 &  05 26 04 & $-$67 29 56 & WC4+O       & 05 & 11.44 & 0.20 & P1031302  \\
SK -67$^{\circ}$106 &  05 26 15 & $-$67 29 58 & B0I         & 16 & 11.78 & 0.15 & A1110101  \\
SK -67$^{\circ}$107 &  05 26 20 & $-$67 29 55 & B0I         & 17 & 12.50 & 0.12 & A1110201  \\
SK -67$^{\circ}$111 &  05 26 48 & $-$67 29 33 & O6:Iafpe    & 15 & 12.57 & 0.11 & P1173001  \\
SK -67$^{\circ}$166 &  05 31 44 & $-$67 38 00 & O4If+       & 10 & 12.27 & 0.10 & A1330100  \\
SK -67$^{\circ}$167 &  05 31 51 & $-$67 39 41 & O4Inf+      & 10 & 12.54 & 0.14 & P1171901  \\
SK -67$^{\circ}$169 &  05 31 51 & $-$67 02 22 & B1Ia        & 08 & 12.18 & 0.20 & P1031603  \\
SK -67$^{\circ}$191 &  05 33 34 & $-$67 30 19 & O8V         & 02 & 13.46 & 0.10 & P1173102  \\
SK -68$^{\circ}$03  &  05 52.1  & $-$68 26    & O9I         & 02 & 13.13 & -    & A0490401  \\
SK -68$^{\circ}$41  &  05 05 27 & $-$68 10 02 & B0.5Ia      & 08 & 12.01 & 0.16 & P1174101  \\
SK -68$^{\circ}$52  &  05 07 20 & $-$68 32 09 & B0Ia        & 01 & 11.70 & 0.15 & P1174001  \\
SK -68$^{\circ}$80  &  05 26 30 & $-$68 50 26 & WC4+O       & 05 & 12.40 & 0.20 & P1031402  \\
SK -68$^{\circ}$135 &  05 37 48 & $-$68 55 08 & ON9.7Ia+    & 01 & 11.36 & 0.25 & P1173901  \\
SK -68$^{\circ}$171 &  05 50 22 & $-$68 11 26 & B1Ia        & 08 & 12.02 & 0.10 & A0490801  \\
SK -69$^{\circ}$59  &  05 03 12 & $-$69 01 37 & B0Ia        & 14 & 12.13 & 0.20 & P1031103  \\
SK -69$^{\circ}$104 &  05 18 59 & $-$69 12 54 & O6Ib(f)     & 01 & 12.10 & 0.11 & P1172401  \\
SK -69$^{\circ}$246 &  05 38 53 & $-$69 02 00 & WN6         & 04 & 11.13 & 0.25 & P1031802  \\
SK -70$^{\circ}$60  &  05 04 40 & $-$70 15 34 & O4-O5V:n    & 15 & 13.85 & 0.13 & P1172001  \\
SK -70$^{\circ}$69  &  05 05 18 & $-$70 25 49 & O5V         & 10 & 13.90 & 0.01 & P1172101  \\
SK -70$^{\circ}$85  &  05 17 05 & $-$70 19 23 & B0I         & 12 & 12.30 & 0.15 & A0491301  \\
SK -70$^{\circ}$91  &  00 59 01 & $-$72 10 28 & O6-O6.5V    & 02 & 13.50 & 0.10 & P1172501  \\
SK -70$^{\circ}$115 &  05 48 49 & $-$70 03 57 & O7Ib        & 18 & 12.24 & 0.22 & P1172601  \\
SK -70$^{\circ}$120 &  05 51 20 & $-$70 17 08 & B1Ia        & 08 & 11.59 & 0.14 & A0491002  \\
SK -71$^{\circ}$45  &  05 31 15 & $-$71 04 08 & O4-5III(f)  & 01 & 11.47 & 0.20 & P1031502  \\
\\
\hline
           &           &             &             &    &       &      &           \\
AV 14      &  00 46 32 & $-$73 06 05 & O3-4V+neb   & 19 & 13.77 & 0.13 & P1175301  \\
AV 15      &  00 46 42 & $-$73 24 54 & O6.5II(f)   & 20 & 13.20 & 0.00 & P1150101  \\
AV 26      &  00 47 50 & $-$73 08 20 & O7III+neb   & 19 & 12.55 & 0.11 & P1176001  \\
AV 47      &  00 48 51 & $-$73 25 57 & O8III((f))  & 20 & 13.40 & 0.00 & P1150202  \\
AV 69      &  00 50 17 & $-$72 53 29 & OC7.5III((f)& 20 & 13.40 & 0.00 & P1150303  \\
AV 70      &  00 50 18 & $-$72 38 09 & O9.5Iw      & 21 & 12.39 & 0.15 & A1180203  \\
AV 75      &  00 50 32 & $-$72 52 36 & O5III(f+)   & 20 & 12.80 & 0.00 & P1150404  \\
AV 81      &  00 50 43 & $-$73 27 06 & WN4.5       & 22 & 13.35 & 0.11 & P2170801  \\
AV 83      &  00 50 52 & $-$72 42 14 & O7Iaf+      & 20 & 13.38 & 0.12 & P1176201  \\
AV 95      &  00 51 21 & $-$72 44 12 & O7III((f))  & 20 & 13.90 & 0.00 & P1150505  \\
AV 170     &  00 55 42 & $-$73 17 30 & O9.7III     & 20 & 14.09 & 0.07 & P2170701  \\
AV 207     &  00 58 33 & $-$71 55 46 & O7V         & 19 & 14.37 & 0.10 & P1175901  \\
AV 232     &  00 59 31 & $-$72 10 45 & O7Iaf+      & 01 & 12.36 & 0.15 & P1030201  \\
AV 238     &  00 59 55 & $-$72 13 37 & O9.5III     & 15 & 13.77 & 0.08 & P1176601  \\
AV 242     &  01 00 06 & $-$72 13 56 & B0.7Iaw     & 23 & 12.11 & 0.03 & P1176901  \\
AV 243     &  01 00 06 & $-$72 47 19 & O6V         & 10 & 13.87 & 0.10 & P1175802  \\
AV 264     &  01 01 07 & $-$71 59 58 & B1Ia        & 24 & 12.36 & 0.03 & P1177001  \\
AV 321     &  01 02 57 & $-$72 08 09 & O9Ib*       & 25 & 13.40 & 0.00 & P1150606  \\
AV 327     &  01 03 10 & $-$72 02 13 & O9.5IIIbw   & 15 & 13.25 & 0.09 & P1176401  \\
AV 372     &  01 04 55 & $-$72 46 47 & O9.5Iabw    & 15 & 12.63 & 0.13 & P1176501  \\
AV 378     &  01 05 09 & $-$72 05 35 & O9III       & 25 & 13.90 & 0.00 & P1150707  \\
AV 388     &  01 05 39 & $-$72 29 26 & O4V         & 19 & 14.12 & 0.11 & P1175401  \\
AV 423     &  01 07 40 & $-$72 50 59 & O9.5II(n)   & 15 & 13.28 & 0.11 & P1176701  \\
AV 451     &  01 10 25 & $-$72 23 28 & O9V         & 19 & 14.15 & 0.08 & P2170601  \\
AV 461     &  01 11 25 & $-$72 09 48 & O8V+neb     & 19 & 14.66 & 0.00 & P2170501  \\
AV 469     &  01 12 28 & $-$72 29 28 & O8II        & 19 & 13.20 & 0.09 & P1176301  \\
AV 488     &  01 15 58 & $-$73 21 24 & B0.5Iaw     & 10 & 11.90 & 0.14 & P1176803  \\
HD 5980    &  00 59 26 & $-$72 09 53 & WN3+OB      & 01 & 11.69 & 0.15 & P1030101  \\
NGC 346-1  &  00 59 04 & $-$72 10 24 & O4III(n)(f) & 26 & 12.57 & 0.13 & P1175501  \\
NGC 346-3  &  00 59 01 & $-$72 10 28 & O2III(f*)   & 06 & 13.50 & 0.10 & P1175201  \\
NGC 346-4  &  00 59 00 & $-$72 10 37 & O5-6V       & 27 & 13.66 & 0.10 & P1175701  \\
NGC 346-6  &  00 58 57 & $-$72 10 33 & O4V((f))    & 28 & 14.02 & 0.09 & P1175601  \\
SK 82      &  00 59 45 & $-$72 44 56 & B0Iaw       & 01 & 12.17 & 0.15 & P1030301  \\
SK 159     &  01 15 59 & $-$73 21 24 & B0.5 Iaw    & 23 & 11.90 & 0.15 & P1030501  \\
\enddata

\tablerefs{1.~Walborn (1977); 2.~Conti et~al. (1986); 3.~Plante (1998);
4.~Smith et~al. (1996); 5.~Smith et~al. (1990); 6.~Walborn et~al. (2002b);
7.~Crowther \& Smith (1997); 8.~Fitzpatrick (1988); 9.~Walborn \& Blades (1997);
10.~Walborn et~al. (1995); 11.~Massey et~al. (1995); 12.~Jaxon et~al. (2001);
13.~Garmany \& Walborn (1987); 14.~Rousseau et~al. (1978);
15.~N.R. Walborn, Private Communication; 16.~Fehrenbach \& Duflot (1982);
17.~PI of the FUSE proposal; 18.~Smith Neubig \& Bruhweiler (1999);
19.~Garmany \& Conti (1985); 20.~Walborn et~al. (2000); 21.~Humphreys (1983);
22.~Breysacher \& Westerlund (1978); 23. Walborn (1983); 24.~Lennon (1997);
25.~Smith Neubig \& Bruhweiler (1997); 26.~Walborn (1978);
27.~Walborn \& Blades (1986); 28.~Massey et~al. (1989).}

\end{deluxetable}

\begin{deluxetable}{ll|ll|ll}
\renewcommand{\arraystretch}{0.85}
\tabletypesize{\scriptsize}
\tablecolumns{6}
\tablewidth{0pc}
\tablecaption{Stars Used for the Solar-Metallicity Library}
\tablehead{ \colhead{Group} & \colhead{Star} & \colhead{Group} & \colhead{Star}
& \colhead{Group} & \colhead{Star}}
\startdata
O3V...............& HD 93205             &B0IV...............  & HD 47417(LiF)     & B1II................& HD 30677     \\
                  & HD 93250             &                     & HD 191495         &                     & HD 75309     \\
                  & HDE 303308            &                     & HD 216044         &                     & HD 169673    \\
                  &                      &                 & BD +53$^{\circ}$2820  &                     & HD 173502    \\
O4V...............& HD 96715             &                     &                   &                     & HD 215733    \\
                  & HD 168076            & B0.5IV............  & HD 97913          &                     & HDE 232522   \\
                  &                      &                     & HD 118571         &                     &              \\
O5V...............& HD 46150             &                     & HD 203938(LiF)    & B1.5II............  & HD 88115     \\
                  & HD 93204             &                     & HD 224257         &                     & HD 12740(LiF)\\
                  &                      &                     &                   &                     &              \\
O6V...............& HD 63005             & B1IV............... & HD 164906         &B2 to B3II.....&=III+I\tablenotemark{a}\\
                  & HD 101131            &                     &                   &                     &              \\
                  & HD 101190            & B2IV............... & HD 110432         & O3I................ & HD 93129A    \\
                  & HD 101298            &                     & HD 116538         &                     &              \\
                  & HD 199579            &                     & HD 125924         & O4I................ & HD 190429    \\
                  & CPD -59$^{\circ}$2600 &                     &                   &                     &              \\
                  &                      & B3IV............... & HD 72088          & O6 \& O6.5I...      & HD 210839    \\
O6.5V............ & HD 5005A             &                     & HD 239683         &                     & HD 163758    \\
                  & HD 39680             &                     & HDE 235874        &                     &              \\
                  & HD 42088(LiF)        &                     &                   & O7I................ & HD 152248    \\
                  & HD 93146             & O3 to O4.5III..& =I+V\tablenotemark{a}  &                     & HD 192639    \\
                  & HD 165052            &                     &                   &                     &              \\
                  &                      & O5III...............& HD 93843          & O8 \& O8.5I...      & HD 74194     \\
O7V...............& HD 91824             &                     &                   &                     & HD 96917     \\
                  & HD 152623            & O6III...............& HD 152233         &                     & HD 175754    \\
                  & CPD -59$^{\circ}$2603 &                     &                   &                     & HD 167971    \\
                  &                      & O6.5III............ & HD 152723         &                     &              \\
O8V...............& HD 13268             &                     & HD 157857         & O9I................ & HD 45314(LiF)\\
                  & HD 66788             &                     & HD 175876         &                     & HD 61347     \\
                  & HD 101413            &                     &                   &                     & HD 210809    \\
                  & HD 165246            & O7III...............& HD 93222          &                     &              \\
                  &                      &                     & HD 167771         & O9.5I.............  & HD 93206     \\
O8.5V             & HD100213             &                     &                   &                     & HD 154368(LiF)\\
                  &                      & O8 \& 8.5III.....   & HD 74920          &                     & HD 218915    \\
O9V...............& HD 91651             &                     & HD 124979         &                     &              \\
                  & HD 93028             &                     &                   & B0I................ & HD 104705    \\
                  & HD 12323             & O9III...............& HD 116852         &                     & HD 167402    \\
                  & HD 201345            &                     & HD 153426         &                     & HD 178487    \\
                  &                      &                     &                   &                     & HD 224868    \\
O9.5V............ & HDE 308813           & O9.5III............ & HD 15642          &                     &              \\
                  & CPD -74$^{\circ}$1569 &                     & HD 89137          & B0.5I.............  & HD 99857     \\
                  &                      &                     & HD 156292         &                     & HD 100276    \\
B0V...............& HD 97471             &                     & HD 166546         &                     & HD 103779    \\
                  & HD 195965            &                     &                   &                     & HD 158661    \\
                  &                      & B0III...............& HD 99890          &                     & HD 179407    \\
B0.5V............ & HD 3827              &                     & HD 100199         &                     & HD 201638    \\
                  & HD 47360             &                     & HD 116781         &                     &              \\
                  & HD 115071            &                     & HD 166716         & B1I................ & HD 92702     \\
                  & BD +35$^{\circ}$4258 &                     & HD 177989         &                     & HD 93827     \\
                  &                      &                     & HD 192035         &                     & HD 94493     \\
B1V...............& HD 121968            &                & CPD -72$^{\circ}$1184  &                     & HD 119608    \\
                  & HD 165955            &                     &                   &                     & HD 151805    \\
                  & HD 212044            & B0.5III............ & HD 172140         &                     & HD 158243    \\
                  & BD +56$^{\circ}$524(LiF) &                     & HD 187459         &                     & HD 160993    \\
                  & CPD -69$^{\circ}$1743 &                     & HD 195455         &                     & HD 163522    \\
                  & LS 277(LiF)          &                     &                   &                     & HD 191877    \\
                  &                      & B1III...............& HD 74711          &                     & HDE 235783   \\
B1.5V............ & HD 118969            &                     & HD 91597          &                     & HDE 332407   \\
                  & HD 121800            &                     & HD 102552         &                     &              \\
                  & HD 146813            &                     & HD 225757         & B2 to B3I.......    & HD 93840     \\
                  &                      &                     &                   &                     &              \\
B2V...............& HD 42401(SiC)        & B2III...............& HD 22586          &                     &              \\
                  & HD 114441            &                     & HD 114444         & WNE.............    & HD 104994    \\
                  & HD 134411            &                     & HD 183899         &                     & HD 187282    \\
                  & HDE 233622           &                     &                   &                     & HD 4004      \\
                  & HDE 315021           & B3III...............& HD 72088          &                     & HD 193077    \\
                  & JL 212               &                     & HD 239683         &                     &              \\
                  &                      &                     & HDE 235874        & WNL.............    & HD 191765    \\
B3V...............& HD 27778             &                     &                   &                     & HD 151932    \\
                  & HD 74662             &O3 to O8II......&=III+I\tablenotemark{a} &                     & HD 96548     \\
                  & HD 179406            &                     &                   &                     &              \\
                  & HD 210121            & O9II................& HD 90087          & WCE.............    & HD 115473    \\
                  & BD +38$^{\circ}$2182 &                     &                   &                     & HD 165763    \\
                  &                      & O9.5II............. & HD 13745          &                     & HD 92809     \\
O3 to O8IV.... & =V+III\tablenotemark{a} &                     & HD 15137          &                     &              \\
                  &                      &                     & HD 168941         & WCL.............    & HD 156385    \\
O9IV..............& HD 60369             &                     &                   &                     & HD 164270    \\
                  &                      & B0.5II............. & HD 168080         &                     & HD 192103    \\
O9.5IV........... & HD 152218            &                     & HD 224151         &                     &              \\
                  & HD 164816            &                     &                   &                     &              \\
\enddata

\tablenotetext{a}{This group was interpolated between the quoted luminosity classes.}

\end{deluxetable}

\begin{deluxetable}{llll|llll}
\tabletypesize{\scriptsize}
\tablecolumns{8}
\tablewidth{0pc}
\tablecaption{Stars Used for the Magellanic Clouds Libraries}
\tablehead{ \colhead{Group} & \colhead{LMC+SMC} & \colhead{LMC} & \colhead{SMC} &
\colhead{Group} & \colhead{LMC+SMC} & \colhead{LMC} & \colhead{SMC}}
\startdata
O3V................& AV 14     & =Solar+SMC\tablenotemark{a}& AV 14    &O8.5II.............&           &    & AV 469  \\
                   &           &               & AV 461                &                   &           &    &         \\
                   &           &               &           &O8 \& O9II......   &           & SK -67$^{\circ}$101 &     \\
O4V................& NGC 346-6 &               & NGC 346-6 &                   &           & SK -68$^{\circ}$03  &      \\
                   & SK -70$^{\circ}$69 &      &                       &                     &           &  &           \\
                   &           &               &                       & O9.5II............. &           &  & AV 327    \\
O4 \& O5V......    &           & SK -70$^{\circ}$69 &                  &                     &           &  & AV 423    \\
                   &           & SK -70$^{\circ}$60 &                  &                     &           &  &           \\
                   &           &               &     & B0II................& =III+I\tablenotemark{b} & =O9II + & =O9.5II \\
O5V................& NGC 346-4 &               & NGC 346-4 &           &        & SK -67$^{\circ}$46\tablenotemark{c} & \\
                   & SK -70$^{\circ}$60 &      &                       &                     &           &  &           \\
                   &           &               &           & O3I.................& =03III    & MK 42         & =O3III \\
O6 \& O6.5V...     &           & SK -66$^{\circ}$18 &                  &                     &           &  &         \\
                   &           & SK -70$^{\circ}$91 & & O4I.................& SK -67$^{\circ}$166 & SK -67$^{\circ}$166 &\\
                   &           &               &      &                     & SK -67$^{\circ}$167 & SK -67$^{\circ}$167 &\\
O6 \& O7V......    &           &               & AV 207                &                     &           &  &           \\
                   &           &               & AV 243                & O4 to O5.5I...      &           &  & =LMC      \\
                   &           &               & AV 388                &                     &           &  &           \\
                   &           &               &      & O6I.................& SK -65$^{\circ}$22  & SK -65$^{\circ}$22  &\\
O6.5V............. & SK -70$^{\circ}$91 &      &      &                     & SK -67$^{\circ}$111 & SK -67$^{\circ}$111 &\\
                   &           &               &      &                     & SK -69$^{\circ}$104 & SK -69$^{\circ}$104 &\\
O7V................& AV 207    & BI 208        &                       &                     &           &  &           \\
                   & BI 208    &               &      & O7I.................& AV 83     & SK -70$^{\circ}$115 & AV 83   \\
                   &           &               &                       &                     & AV 232    &  & AV 232    \\
O9V............... & AV 451    & =Solar+SMC\tablenotemark{a} &AV 451   &                     &           &  &           \\
                   &           &               &                       & O9 \& O9.5I...      & AV 321    & BI 170 & AV 70 \\
B0V................& =O9V      & =Solar+SMC\tablenotemark{a} & =O9V    &   & AV 372    & SK -65$^{\circ}$21 & AV 372    \\
                   &           &               &                       &   & BI 170    & SK -66$^{\circ}$169 &          \\
O3-B0IV.........&=V+III\tablenotemark{b}&=V+III\tablenotemark{b}&=V+III\tablenotemark{b}&&SK -65$^{\circ}$21&SK -67$^{\circ}$ 5&\\
                   &           &               &                       &   & SK -66$^{\circ}$169& SK -68$^{\circ}$135 & \\
O3III..............& NGC 346-3 & HD 269810     & NGC 346-3             &   & SK -67$^{\circ}$05 &      &           \\
                   & HD 269810 & SK -66$^{\circ}$172 &                 &   & SK -68$^{\circ}$135&      &           \\
                   & SK -66$^{\circ}$172  &    &                       &                     &           &  &           \\
                   &           &               &      & B0I.................& AV 242    & SK -66$^{\circ}$185 & AV 242   \\
O4III..............&SK -71$^{\circ}$45 &SK -71$^{\circ}$45 &NGC 346-1  &   & SK 82     & SK -67$^{\circ}$14  & AV 264   \\
                   & SK -67$^{\circ}$69 & SK -67$^{\circ}$69 &         &   & SK 159    & SK -67$^{\circ}$28  & AV 488    \\
                   &           &               &                   & & SK -67$^{\circ}$14 & SK -67$^{\circ}$76  & SK 82  \\
O5III..............& AV 75     &               & AV 75             & & SK -67$^{\circ}$76 & SK -67$^{\circ}$106 & SK 159 \\
                   &           &               &         &         & SK -67$^{\circ}$169& SK -67$^{\circ}$107 & \\
O6III..............& SK -66$^{\circ}$100&      &         &         & SK -68$^{\circ}$41 & SK -67$^{\circ}$169 & \\
                   &           &               &         &         & SK -68$^{\circ}$52 & SK -68$^{\circ}$52  & \\
O7III..............& AV 69     & BI 229(SiC)   & AV 26   &         & SK -69$^{\circ}$59 & SK -68$^{\circ}$41  & \\
                   & AV 95     & BI 272(LiF)   & AV 69   &         &           & SK -68$^{\circ}$171     &      \\
                   & BI 229    &               & AV 95   &         &           & SK -69$^{\circ}$59      &      \\
                   &           &               &         &         &           & SK -70$^{\circ}$85(LiF) &      \\
O8III..............& AV 47     & SK -67$^{\circ}$191 & AV 47       &           &           & SK -70$^{\circ}$120 & \\
                   & AV 378    & BI 173        & AV 378            &                     &           &  &          \\
                   & SK -67$^{\circ}$191 &     &                       &                     &           &  &      \\
                   &           &               &         & WNE...............  & HD 5980   & HD 32109      & HD 5980   \\
O9III..............& AV 238    &               &         &             & HD 32109  & SK -69$^{\circ}$246    & AV 81 \\
                   &           &               &         &                     & SK -69$^{\circ}$246 &     &           \\
O9 \& O9.5III..    & &=Solar+SMC\tablenotemark{a} & AV 238             &                     &           &  &           \\
                   &           &               & AV 170  & WNL...............  & HD 33133  & HD 33133      & =LMC       \\
                   &           &               &         &                     & HDE 269927& HDE 269927    &           \\
B0III..............& =O9III & =Solar+SMC\tablenotemark{a} & =O9.5III   &       & HDE 269582& HDE 269582    &           \\
                   &           &               &         &                     & HDE 269687& HDE 269687    &           \\
O3 to 6II........  &=III+I\tablenotemark{b}&&=III+I\tablenotemark{b}   &                     &           &  &           \\
                   &           &               &         & WCE...............  & HD 37026  & HD 37026      & =LMC       \\
O3 to O7II.....    &           & =III+I\tablenotemark{b} &             & & SK -67$^{\circ}$104& SK -67$^{\circ}$104 & \\
                   &           &               &         &             & SK -68$^{\circ}$80 & SK -68$^{\circ}$80  & \\
O6.5II.............&           &               & AV 15                 &                     &           &  &           \\
                   &           &               &         & WCL...............  & =Solar    & =Solar        & =Solar     \\
O7.5 \& O8II...    & AV 15     &               &                       &                     &           &  &           \\
                   & SK -67$^{\circ}$101 &     &                       &                     &           &  &           \\
\enddata

\tablenotetext{a}{This group was interpolated between the quoted libraries.}
\tablenotetext{b}{This group was interpolated between the quoted luminosity classes.}
\tablenotetext{c}{This group was interpolated between the LMC O9II group and the LMC star SK-67$^\circ$46 (B1.5I).}

\end{deluxetable}

\clearpage

\begin{deluxetable}{ccccccccccc}
\tabletypesize{\scriptsize}
\tablecaption{log $L_{1160}$ (in erg s$^{-1}$ \AA$^{-1}$):
{\it INSTANTANEOUS} starburst}
\tablewidth{0pt}
\tablehead{
\colhead{} &
\colhead{} &
\multicolumn{4}{c}{Z$_{\odot}$} &
\colhead{} &
\multicolumn{4}{c}{$1/4$~Z$_{\odot}$} \\
\cline{3-6}\cline{8-11} \\
\colhead{Age} &
\colhead{} &
\colhead{$\alpha$ = 2.35,} &
\colhead{$\alpha$ = 2.35,} &
\colhead{$\alpha$ = 3.3,} &
\colhead{$\alpha$ = 1.5,} &
\colhead{} &
\colhead{$\alpha$ = 2.35,} &
\colhead{$\alpha$ = 2.35,} &
\colhead{$\alpha$ = 3.3,} &
\colhead{$\alpha$ = 1.5,} \\
\colhead{(Myr)} &
\colhead{} &
\colhead{100 M$_{\odot}$} &
\colhead{30 M$_{\odot}$} &
\colhead{100 M$_{\odot}$} &
\colhead{100 M$_{\odot}$} &
\colhead{} &
\colhead{100 M$_{\odot}$} &
\colhead{30 M$_{\odot}$} &
\colhead{100 M$_{\odot}$} &
\colhead{100 M$_{\odot}$} \\
}
\startdata
0  & & 39.551 & 39.133 & 38.629 & 40.065 & & 39.516 & 39.194 & 38.658 & 39.998 \\
1  & & 39.616 & 39.153 & 38.674 & 40.141 & & 39.566 & 39.204 & 38.690 & 40.057 \\
2  & & 39.696 & 39.194 & 38.726 & 40.238 & & 39.651 & 39.229 & 38.733 & 40.177 \\
3  & & 39.679 & 39.228 & 38.747 & 40.160 & & 39.768 & 39.258 & 38.801 & 40.323 \\
5  & & 39.336 & 39.307 & 38.611 & 39.579 & & 39.490 & 39.322 & 38.720 & 39.802 \\
7  & & 39.096 & 39.157 & 38.488 & 39.217 & & 39.279 & 39.340 & 38.634 & 39.449 \\
10 & & 38.869 & 38.929 & 38.355 & 38.891 & & 39.089 & 39.149 & 38.540 & 39.157 \\
\enddata

{NOTE : For these models, $M_{low}$~=~1~M$_{\odot}$
and $10^6$~M$_\odot$ was converted into stars at the initial time of the
star formation.}

\end{deluxetable}

\begin{deluxetable}{ccccccccccc}
\tabletypesize{\scriptsize}
\tablecaption{log $L_{1160}$ (in erg s$^{-1}$ \AA$^{-1}$):
{\it CONTINUOUS} star formation}
\tablewidth{0pt}
\tablehead{
\colhead{} &
\colhead{} &
\multicolumn{4}{c}{Z$_{\odot}$} &
\colhead{} &
\multicolumn{4}{c}{$1/4$~Z$_{\odot}$} \\
\cline{3-6}\cline{8-11} \\
\colhead{Age} &
\colhead{} &
\colhead{$\alpha$ = 2.35,} &
\colhead{$\alpha$ = 2.35,} &
\colhead{$\alpha$ = 3.3,} &
\colhead{$\alpha$ = 1.5,} &
\colhead{} &
\colhead{$\alpha$ = 2.35,} &
\colhead{$\alpha$ = 2.35,} &
\colhead{$\alpha$ = 3.3,} &
\colhead{$\alpha$ = 1.5,} \\
\colhead{(Myr)} &
\colhead{} &
\colhead{100 M$_{\odot}$} &
\colhead{30 M$_{\odot}$} &
\colhead{100 M$_{\odot}$} &
\colhead{100 M$_{\odot}$} &
\colhead{} &
\colhead{100 M$_{\odot}$} &
\colhead{30 M$_{\odot}$} &
\colhead{100 M$_{\odot}$} &
\colhead{100 M$_{\odot}$} \\
}
\startdata
0   & & 38.551 & 38.133 & 37.629 & 39.065 & & 38.516 & 38.194 & 37.658 & 38.998 \\
10  & & 40.422 & 40.186 & 39.602 & 40.845 & & 40.486 & 40.278 & 39.687 & 40.899 \\
100 & & 40.562 & 40.429 & 39.943 & 40.888 & & 40.678 & 40.586 & 40.123 & 40.963 \\
500 & & 40.570 & 40.443 & 39.974 & 40.889 & & 40.706 & 40.626 & 40.222 & 40.966 \\
\enddata

{NOTE : For these models, $M_{low}$~=~1~M$_{\odot}$
and a star-forming rate of 1~M$_\odot$~yr$^{-1}$ was adopted.}

\end{deluxetable}

\clearpage

\begin{figure} 
\plotone{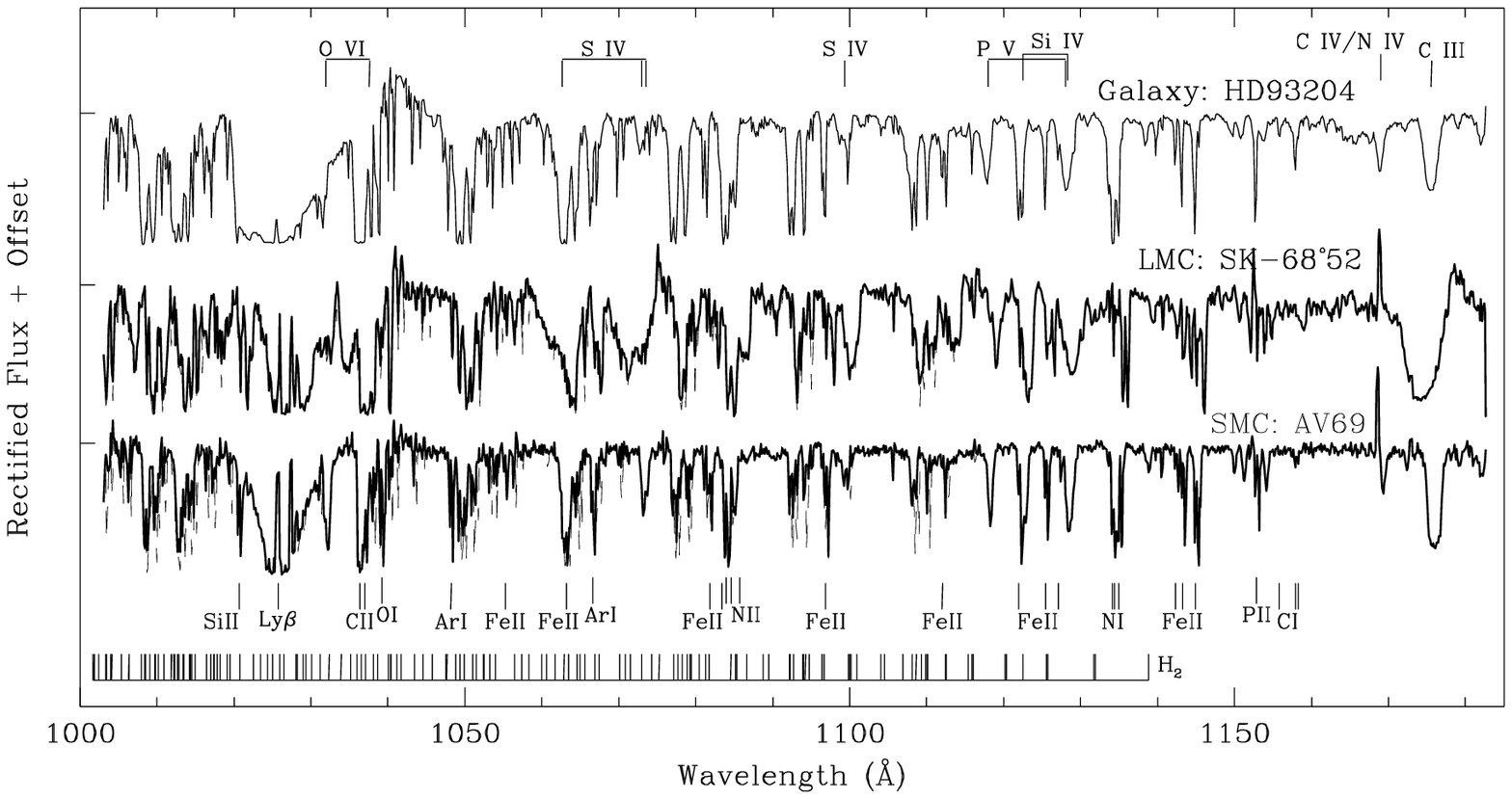} 
\end{figure} 
 
\begin{figure} 
\plotone{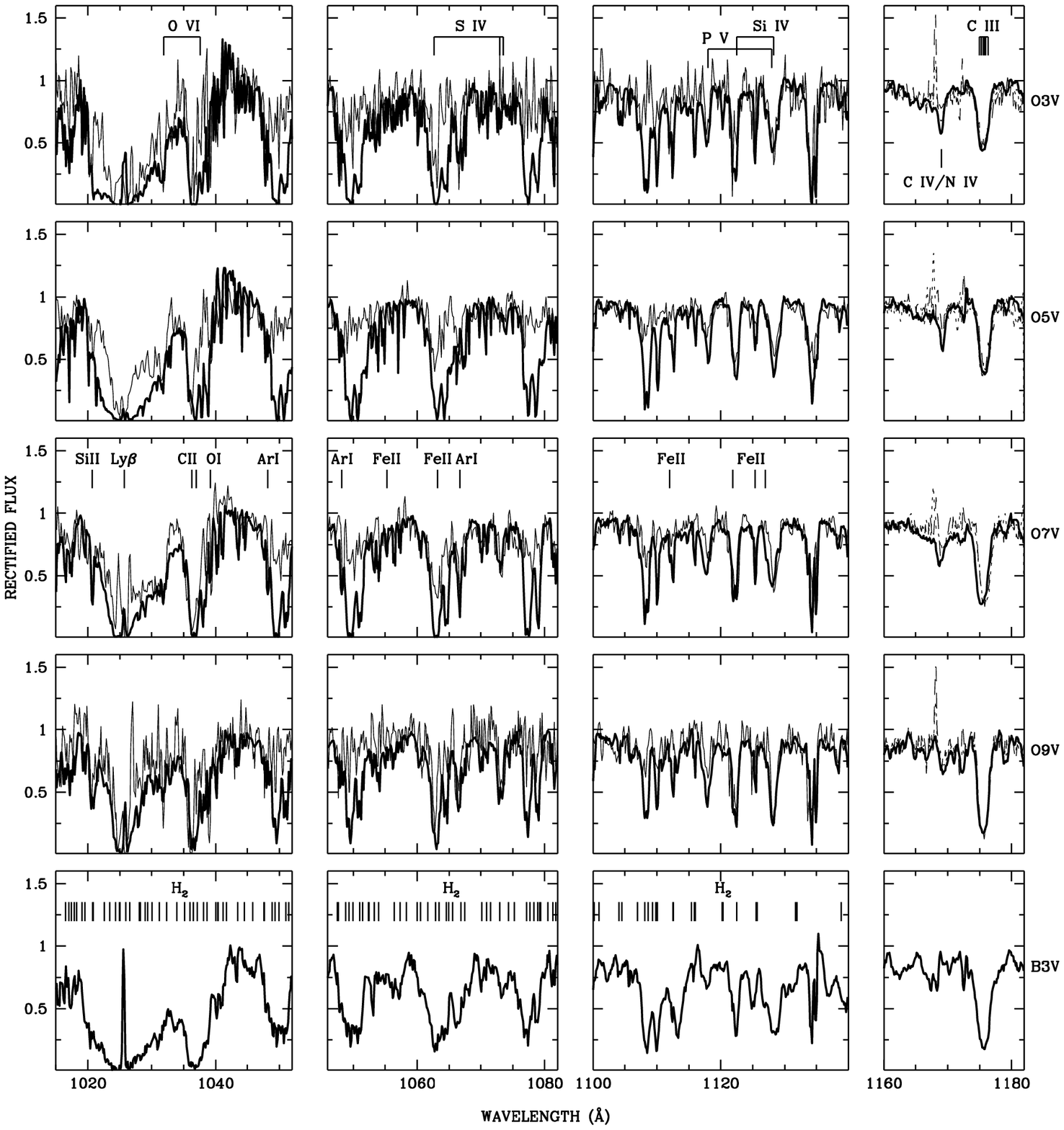} 
\end{figure} 
 
\begin{figure} 
\plotone{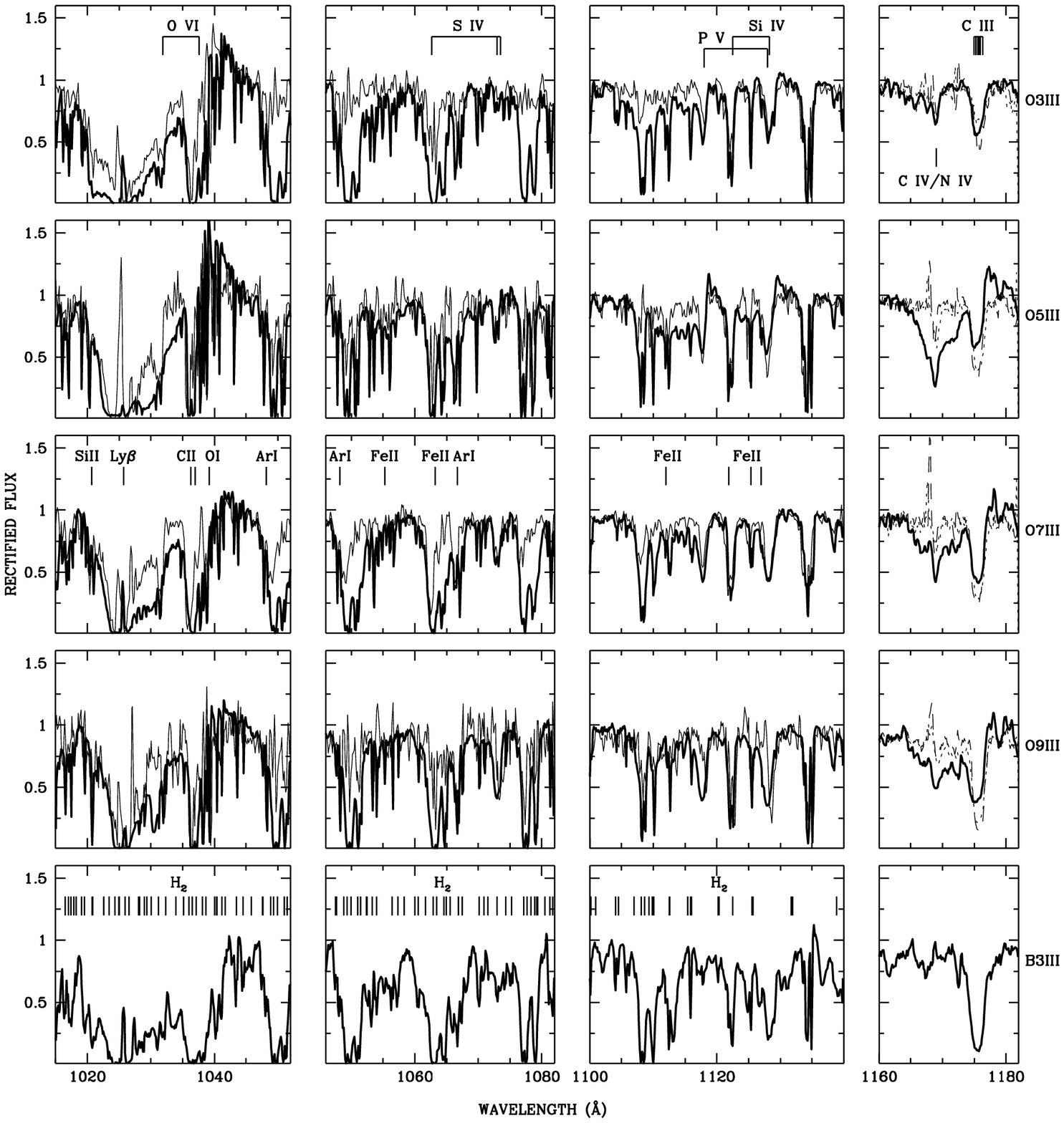} 
\end{figure} 
 
\begin{figure} 
\plotone{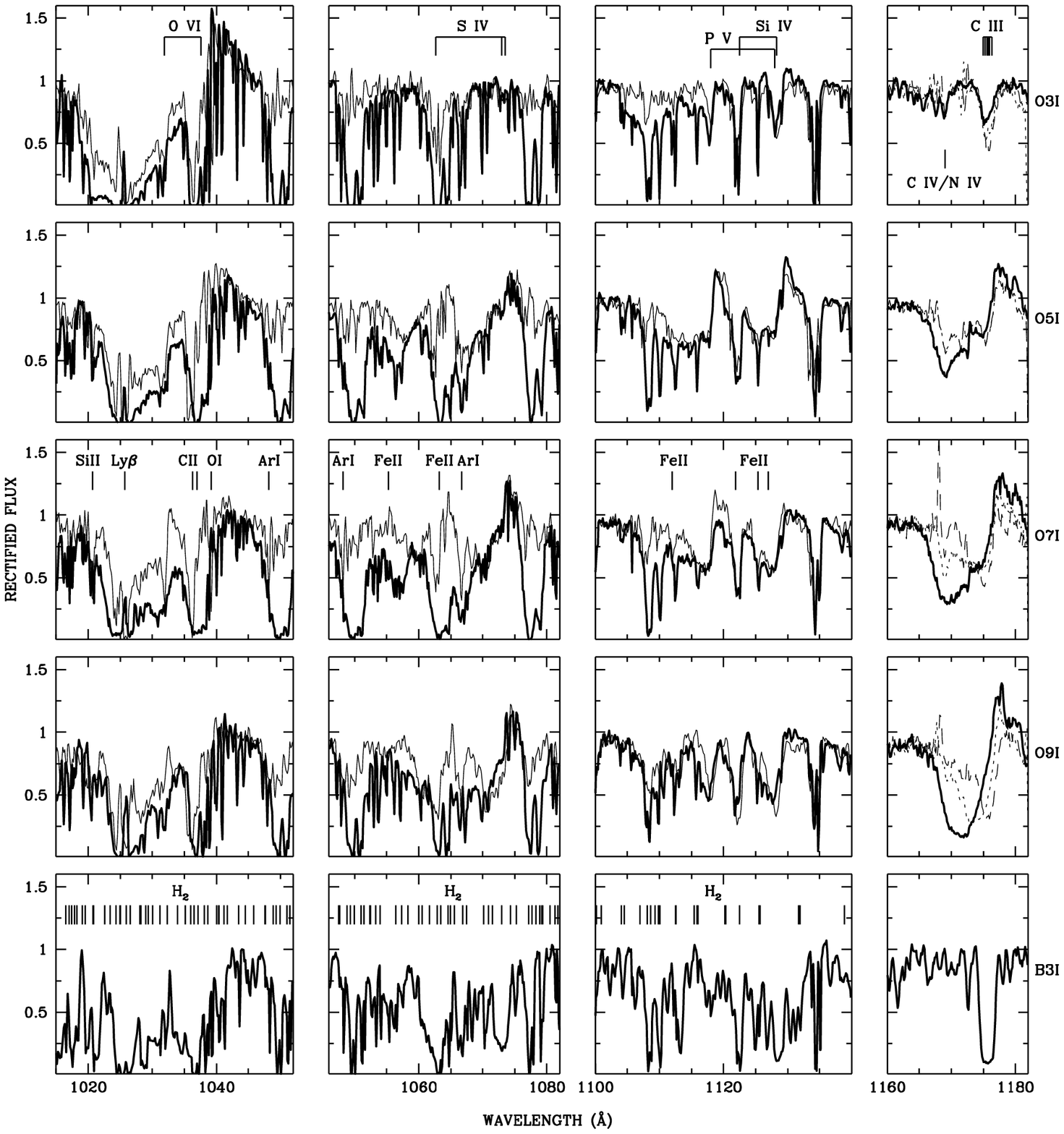} 
\end{figure} 
 
\begin{figure} 
\plotone{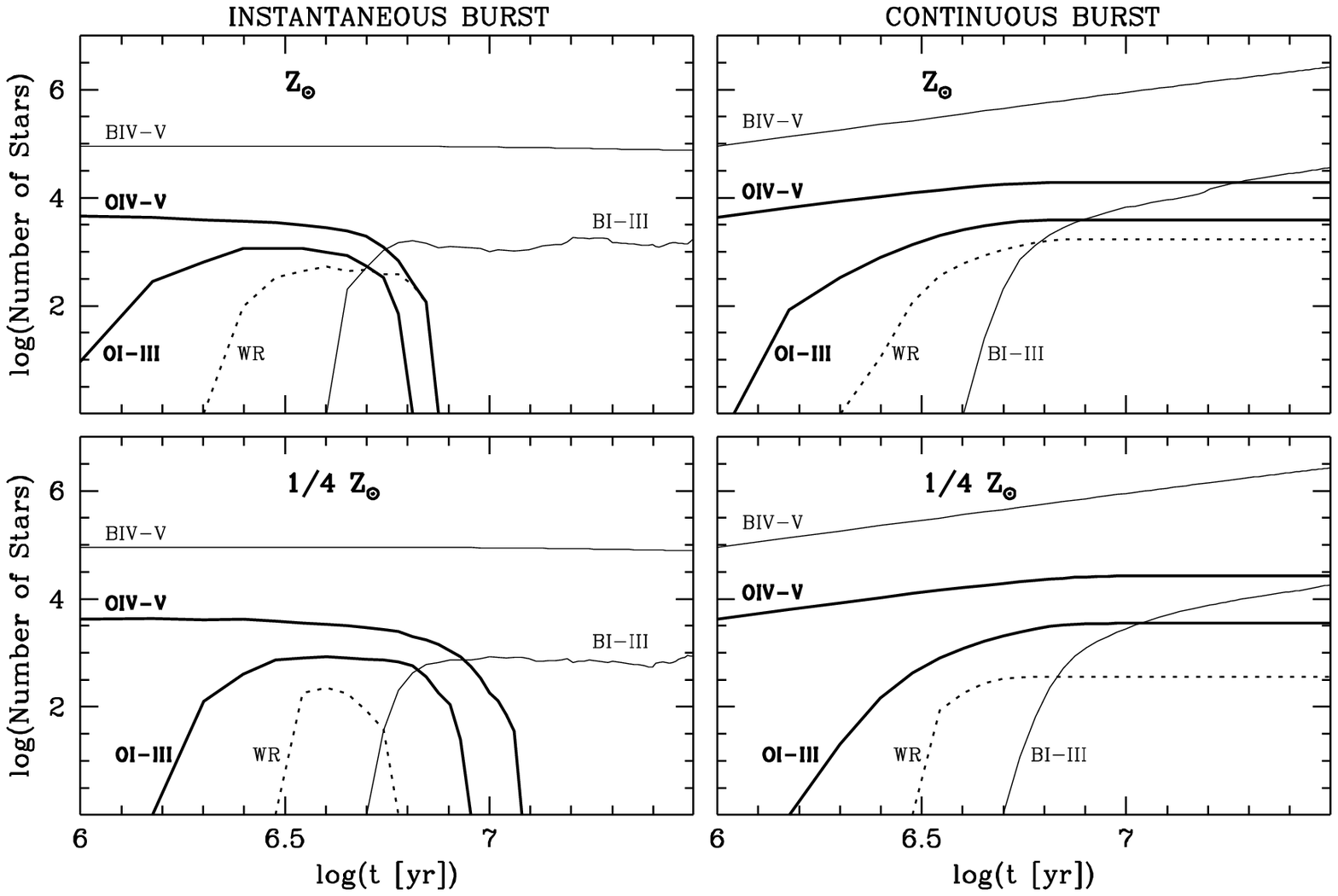} 
\end{figure} 
 
\begin{figure} 
\plotone{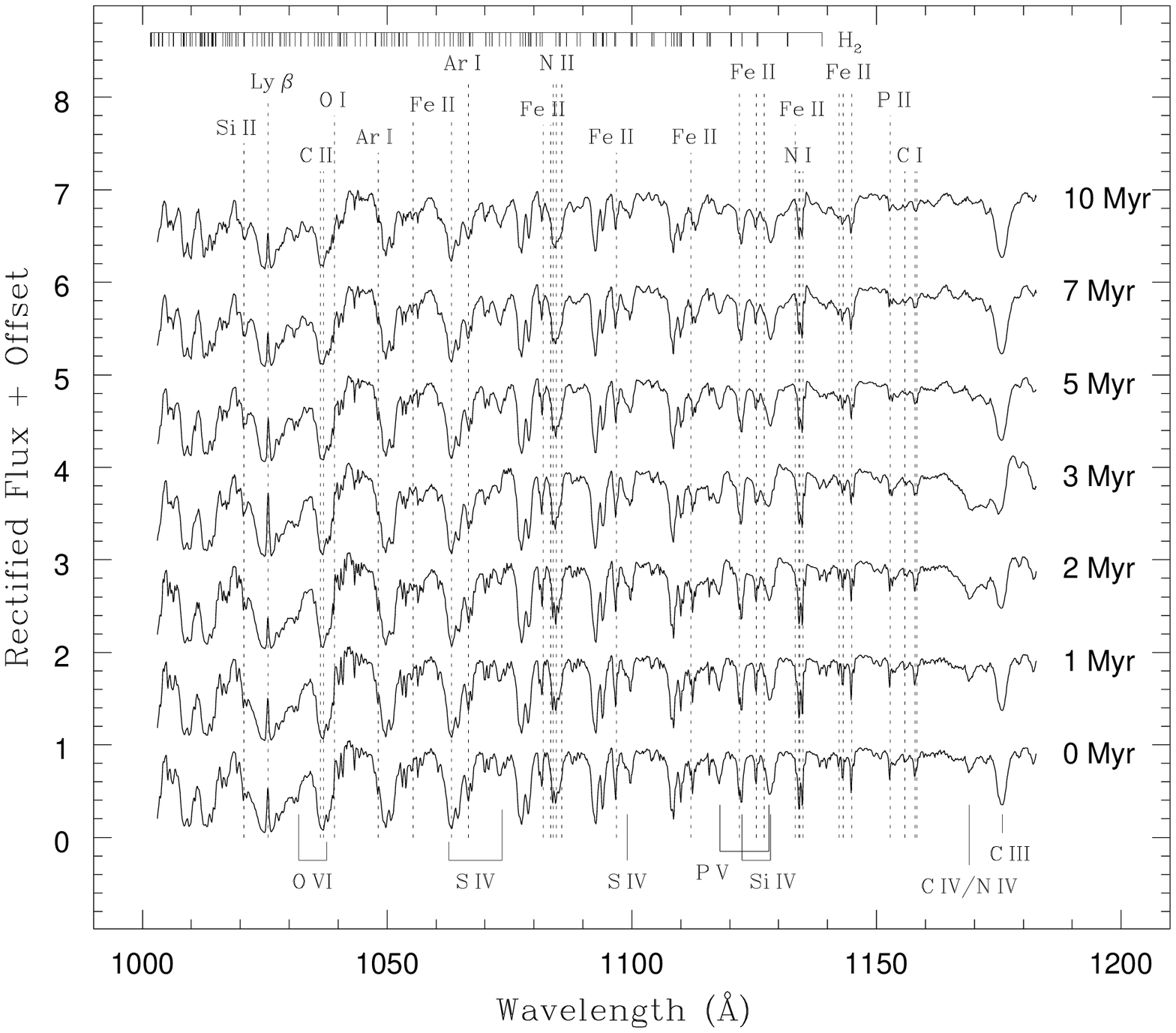} 
\end{figure} 
 
\begin{figure} 
\plotone{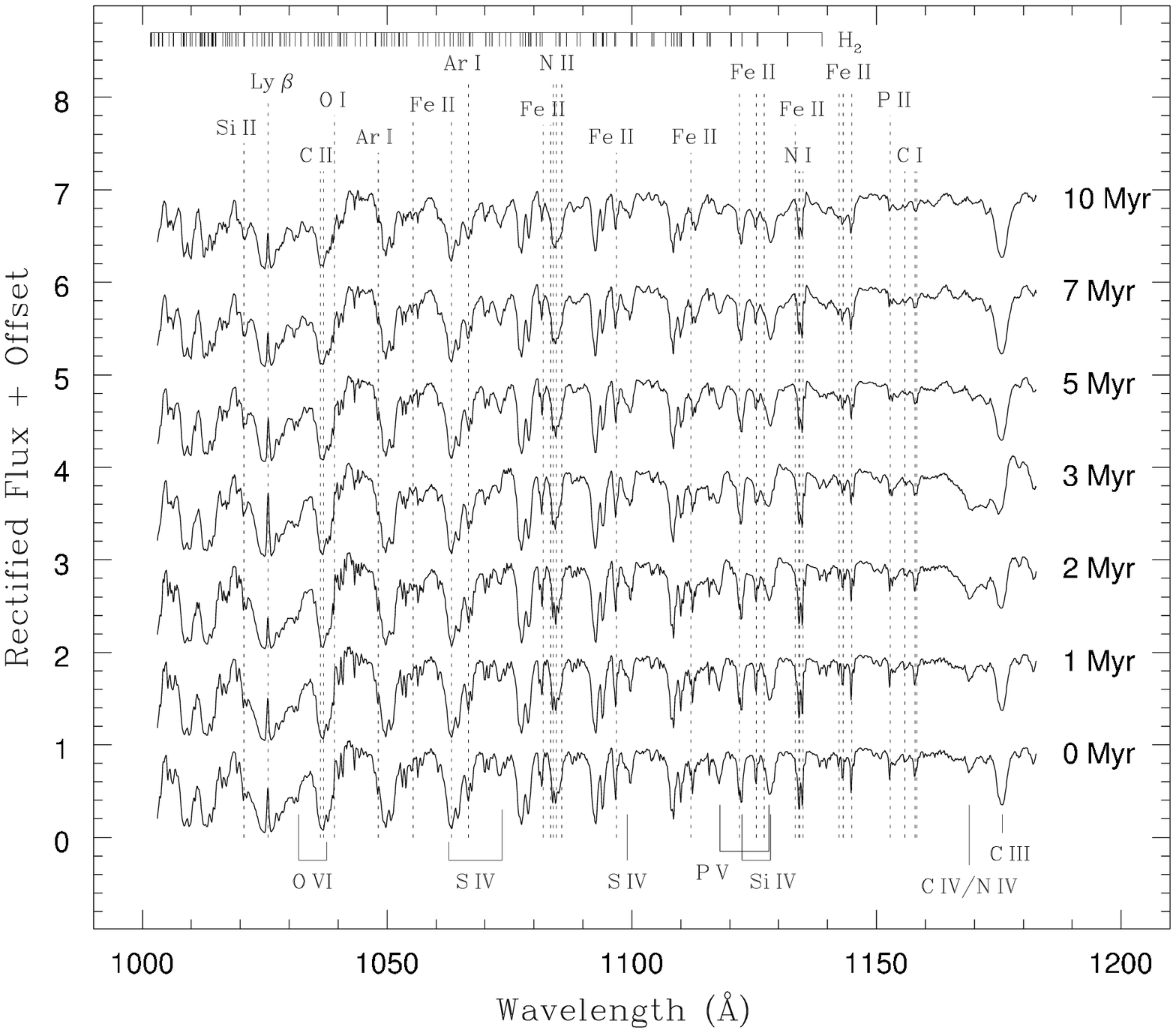} 
\end{figure} 
 
\begin{figure} 
\plotone{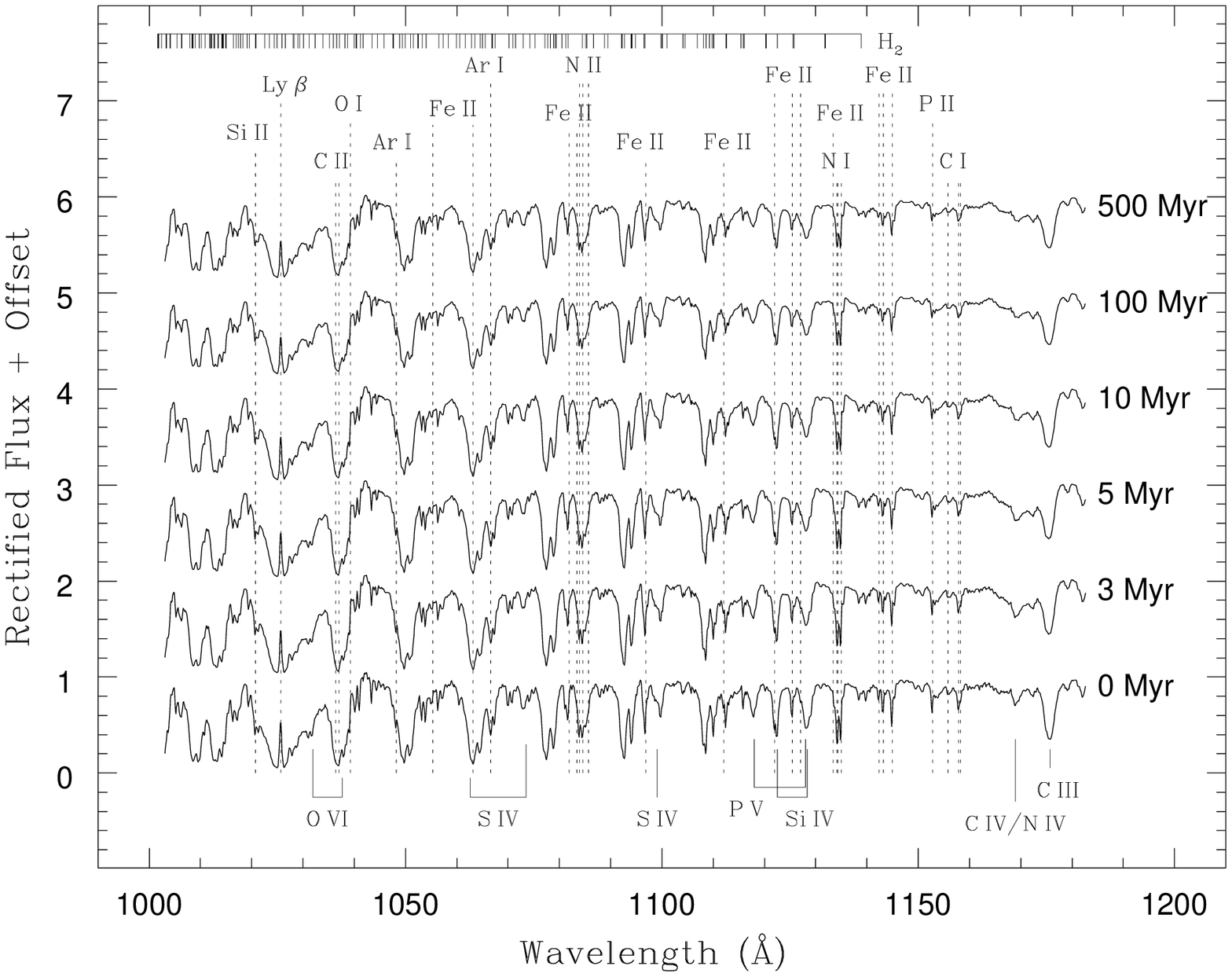} 
\end{figure} 
 
\begin{figure} 
\plotone{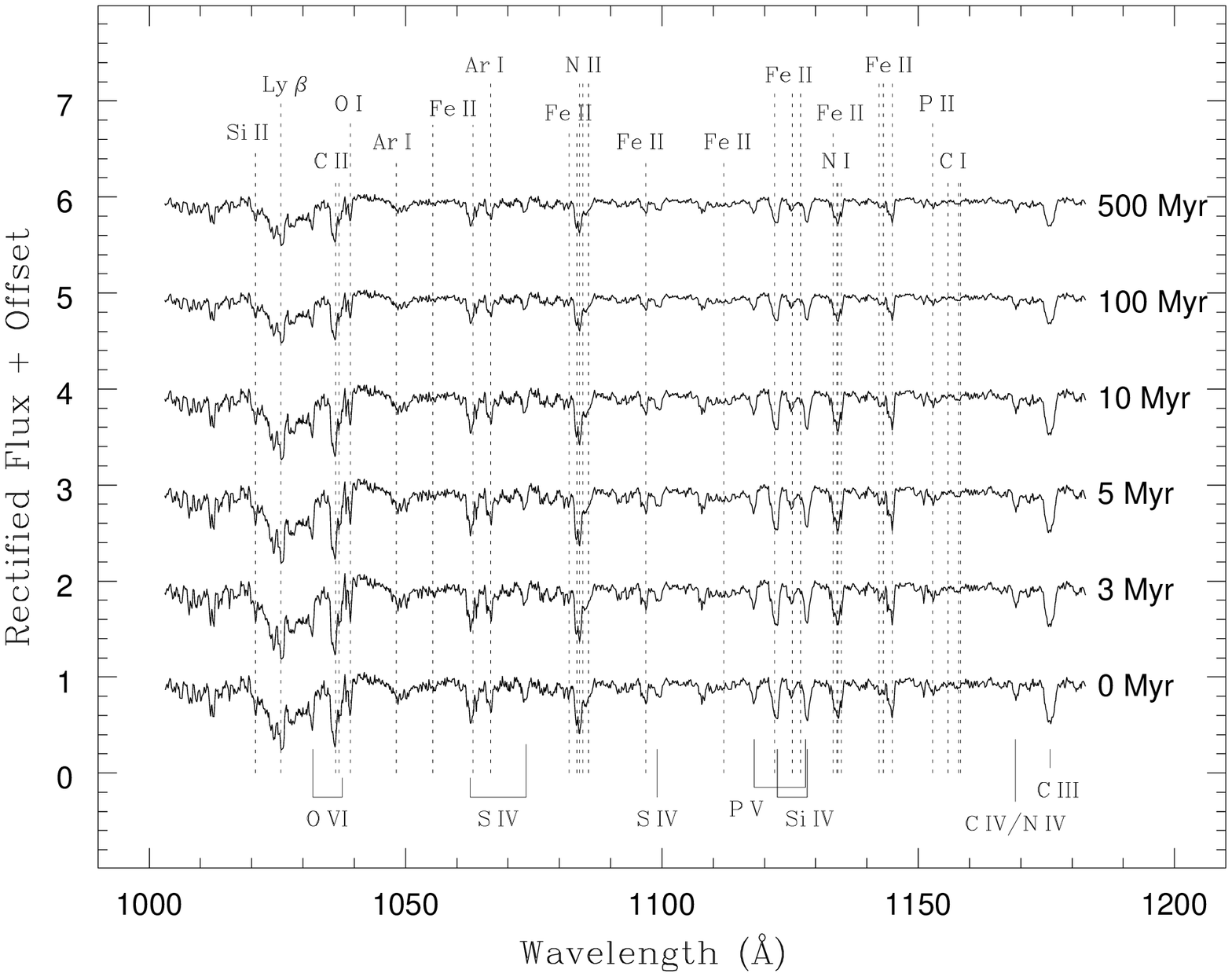} 
\end{figure} 
 
\begin{figure} 
\plotone{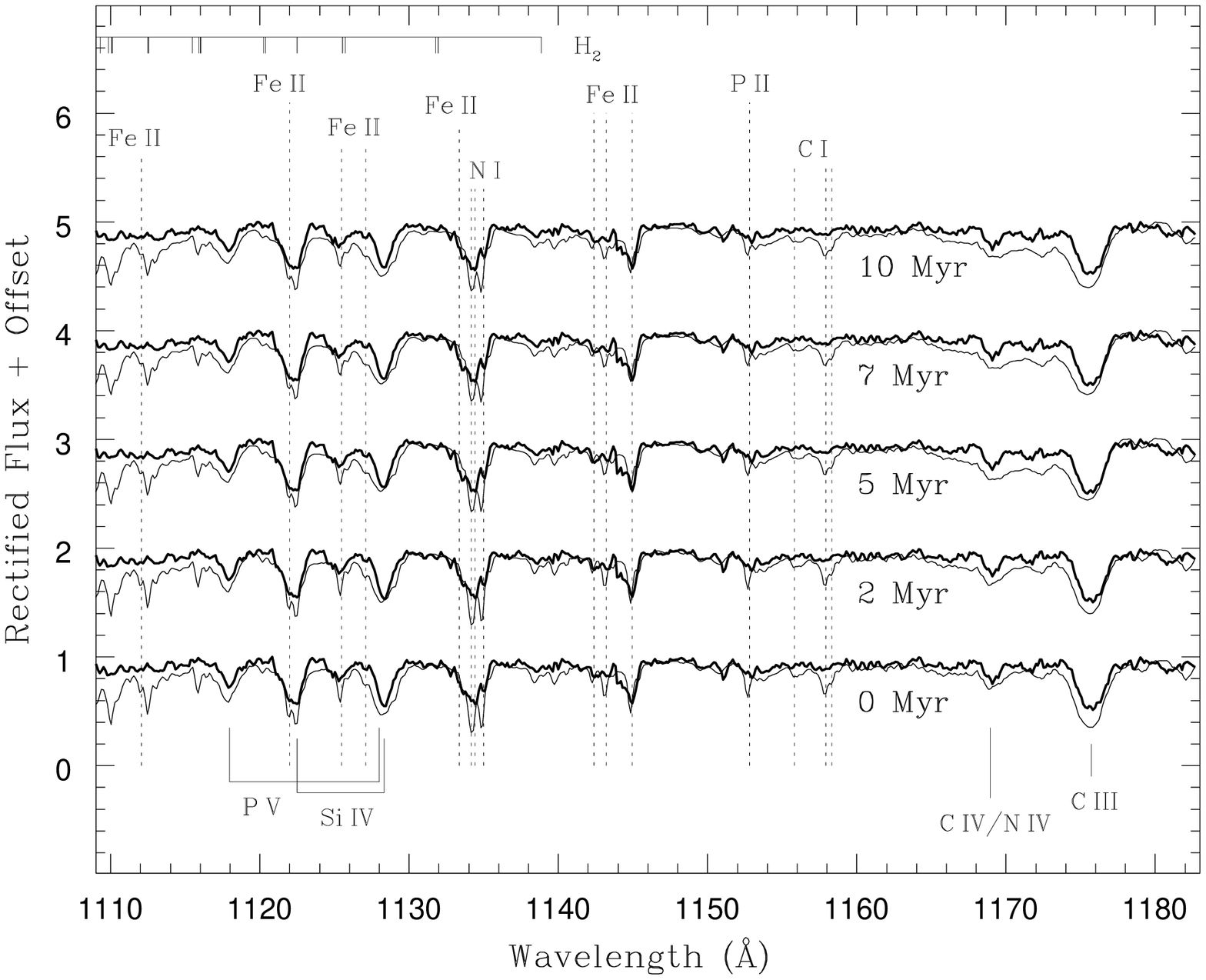} 
\end{figure} 
 
\begin{figure} 
\plotone{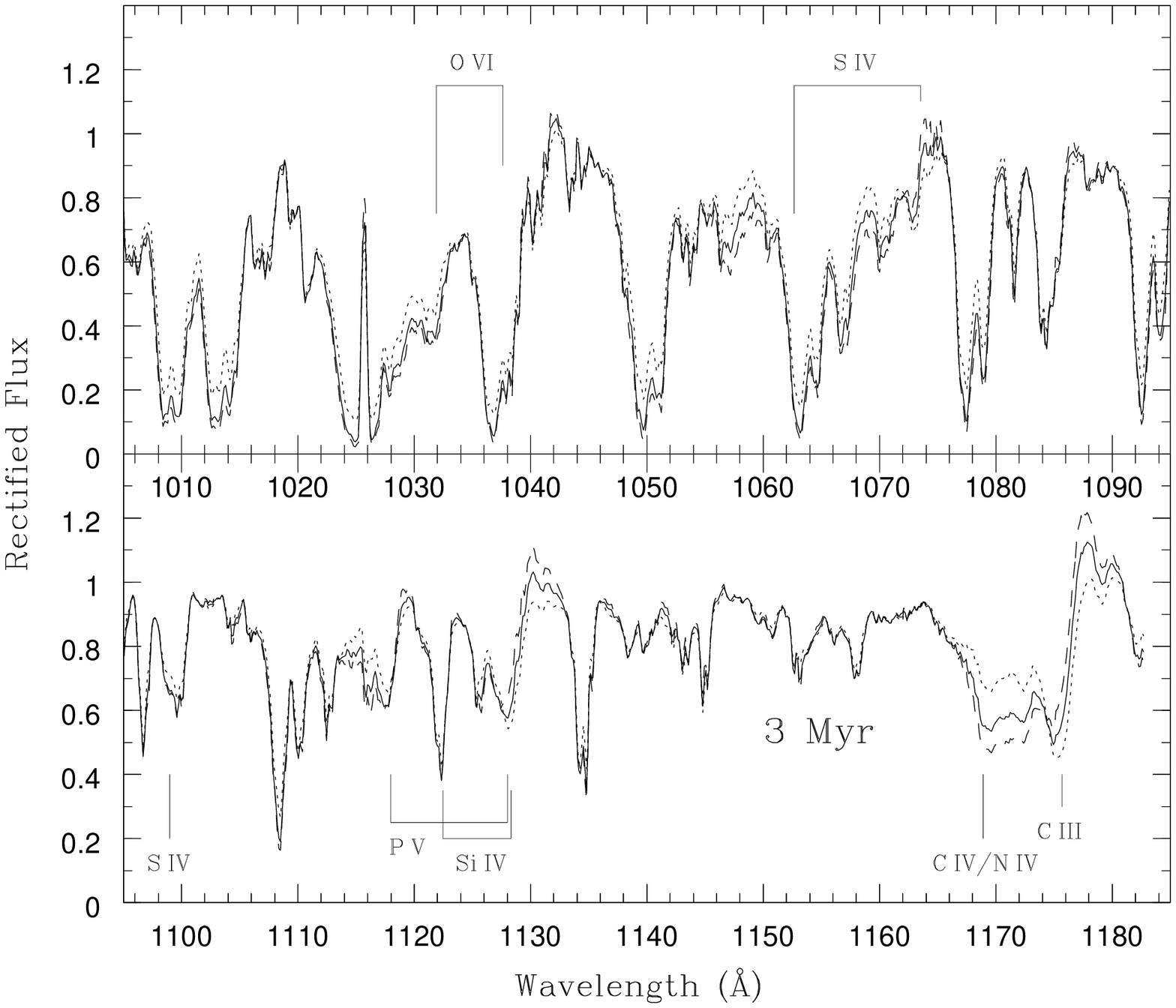} 
\end{figure} 
 
\begin{figure} 
\plotone{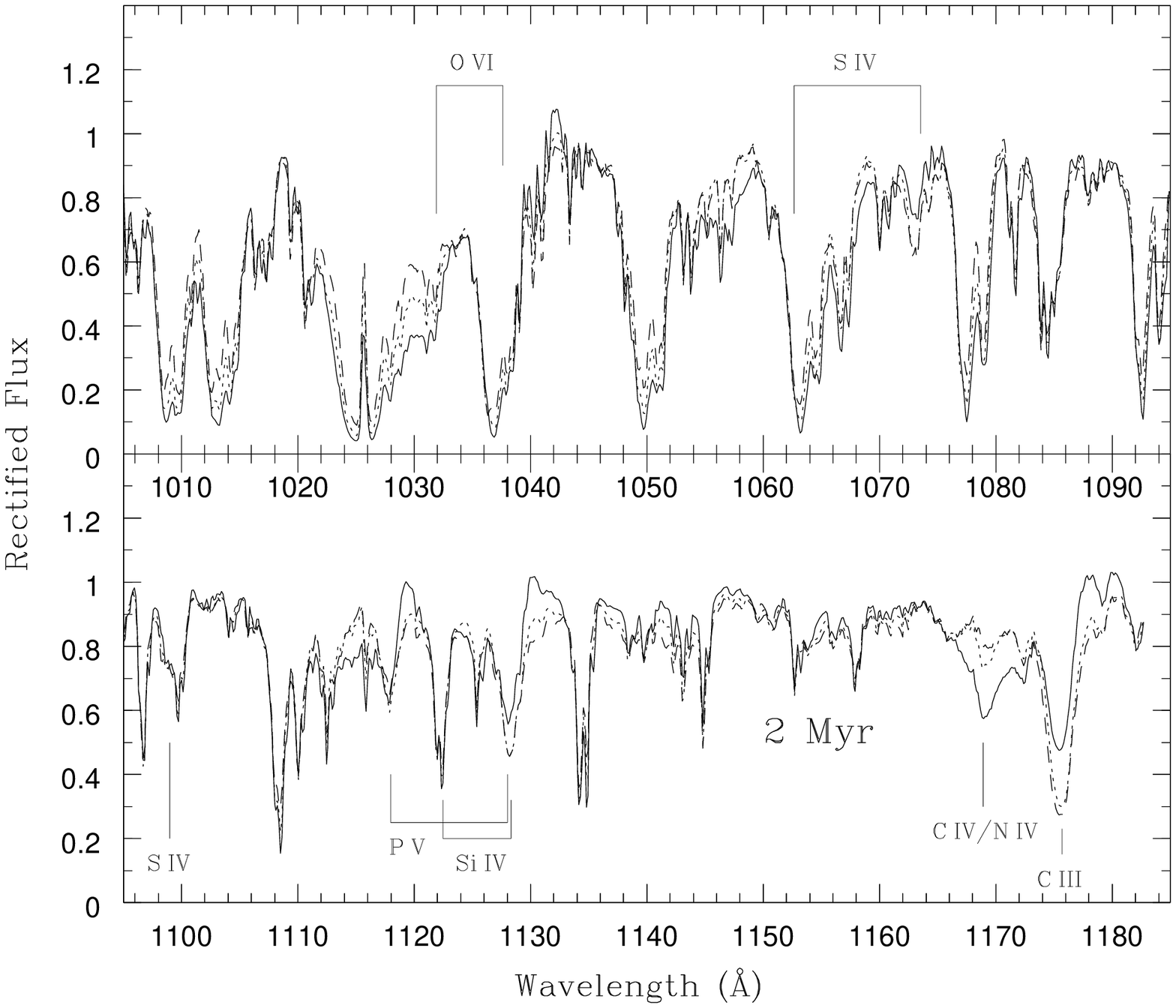} 
\end{figure} 
 
\begin{figure} 
\plotone{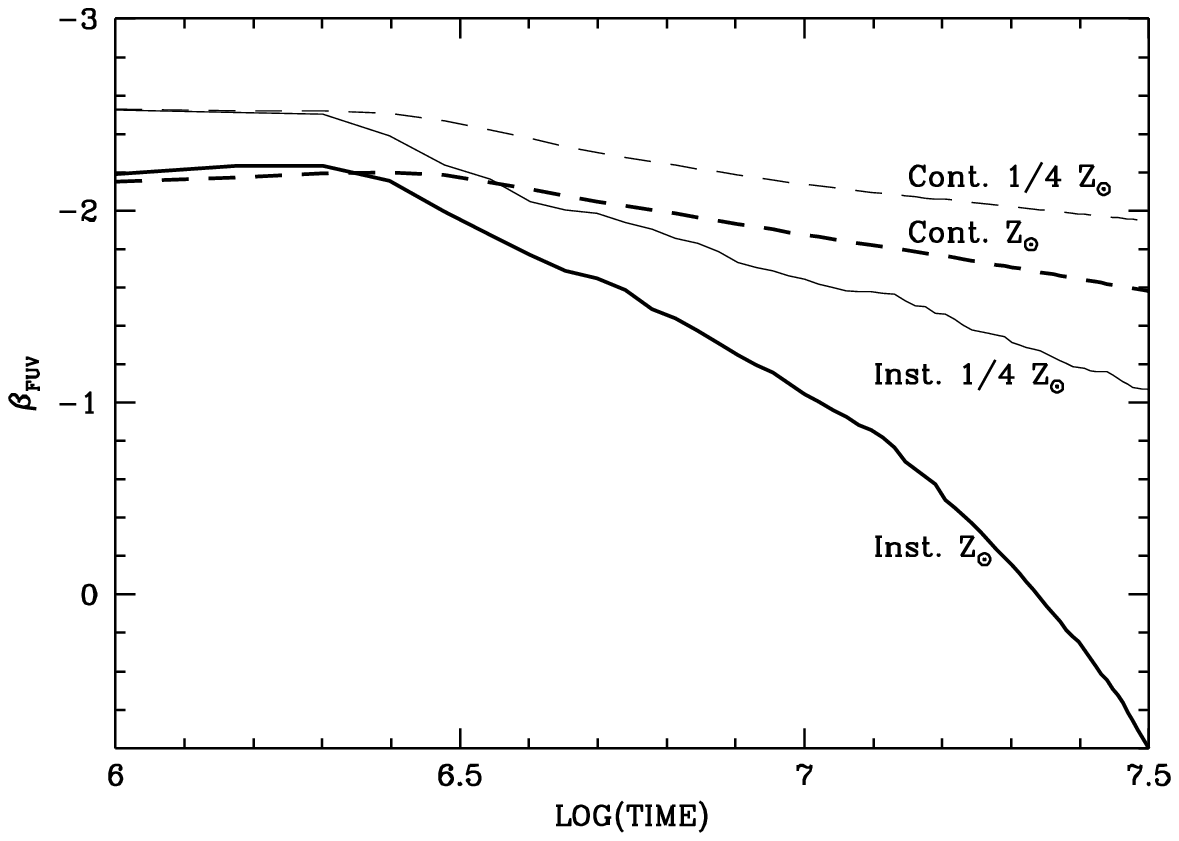} 
\end{figure} 
 
\end{document}